\documentclass{article}

\usepackage[left=2.5cm,right=2.5cm,top=2.5cm,bottom=2.5cm]{geometry}

\usepackage[latin1]{inputenc}
\usepackage[english]{babel} 
\usepackage{amsmath,amsfonts,amsthm,bm} 
\usepackage{graphicx}
\usepackage{textcomp}
\usepackage{gensymb}
\usepackage{dcolumn}
\usepackage{bm}
\usepackage{color}
\usepackage[autostyle]{csquotes}
\usepackage{tikz}
\usepackage{setspace}
\usepackage{multirow}
\usepackage{multicol}

\usepackage{hyperref}

\usepackage{xcolor}

\title{Generalized and multi-oscillation solitons in the \\ Nonlinear Schr{\"o}dinger Equation with quartic dispersion}

\author{Ravindra Bandara\footnotemark[1] \and Andrus Giraldo\footnotemark[2] 
\and Neil G.R. Broderick\footnotemark[3] \and Bernd Krauskopf\footnotemark[4]}

\date{}
\begin{document}
\maketitle

\renewcommand{\thefootnote}{\fnsymbol{footnote}}
\footnotetext[1]{Department of Mathematics and Dodd-Walls Centre for Photonic and Quantum Technologies, The University of Auckland, Private Bag 92019, Auckland 1142, New Zealand
  (\href{mailto:ravindra.bandara@auckland.ac.nz}{ravindra.bandara@auckland.ac.nz})}
\footnotetext[2]{School of Computational Sciences, Korea Institute for Advanced Study, Seoul 02455, Korea
  (\href{mailto:agiraldo@kias.re.kr}{agiraldo@kias.re.kr})}
\footnotetext[3]{Department of Physics and Dodd-Walls Centre for Photonic and Quantum Technologies, The University of Auckland, Private Bag 92019, Auckland 1142, New Zealand
  (\href{mailto:n.broderick@auckland.ac.nz}{n.broderick@auckland.ac.nz})}
\footnotetext[4]{Department of Mathematics and Dodd-Walls Centre for Photonic and Quantum Technologies, The University of Auckland, Private Bag 92019, Auckland 1142, New Zealand
  (\href{mailto:b.krauskopf@auckland.ac.nz}{b.krauskopf@auckland.ac.nz})}
\renewcommand{\thefootnote}{\arabic{footnote}}

\begin{abstract}
We study different types of solitons of a generalized nonlinear Schr{\"o}dinger equation (GNLSE) that models optical pulses traveling down an optical waveguide with quadratic as well as quartic dispersion. A traveling-wave ansatz transforms this partial differential equation into a fourth-order nonlinear ordinary differential equation (ODE) that is Hamiltonian and has two reversible symmetries. Homoclinic orbits of the ODE that connect the origin to itself represent solitons of the GNLSE, and this allows us to study the existence and organization of solitons with advanced numerical tools for the detection and continuation of connecting orbits. In this way, we establish the existence of connections from one periodic orbit to another, called PtoP connections. They give rise to families of homoclinic orbits to either of the two periodic orbits; in the GNLSE they correspond to \emph{generalized solitons} with oscillating tails whose amplitude does not decay but reaches a nonzero limit. Moreover, PtoP connections can be found in the energy level of the origin, where connections between this equilibrium and a given periodic orbit, called EtoP connections, are known to organize families of solitons. As we show here, EtoP and PtoP cycles can be assembled into different types of heteroclinic cycles that give rise to additional families of homoclinic orbits to the origin. In the GNLSE, these correspond to \emph{multi-oscillation solitons} that feature several episodes of different oscillations in between their decaying tails. As for solitons organized by EtoP connections only, multi-oscillation solitons are shown to be an integral part of the phenomenon of truncated homoclinic snaking. 
\end{abstract}

\section{Introduction}
\label{sec:intro}

Optical solitons exist in photonic waveguides, notably, in  optical fibers, due to the balance between dispersion and the Kerr nonlinearity, and they have been the focus of many recent experimental \cite{PhysRevA.87.025801,blanco2016pure}
 and theoretical \cite{tam2018solitary, tam2019stationary, GDK,karlsson1994soliton,AKHMEDIEV1994540,PhysRevA.103.063514} studies. In the first instance, the dispersion has been modeled with a quadratic term. However, it has recently been discovered that waveguides with only quartic dispersion may support solitons with oscillating tails, which are much narrower in a well defined way (in terms of how the amplitude relates to the width \cite{blanco2016pure}). This discovery of quartic solitons has led to considerable interest in characterizing the interplay between quartic and quadratic dispersion and the types of solitons this may entail \cite{AKHMEDIEV1995109,PhysRevA.103.063514,tam2018solitary,tam2019stationary,GDK}. To set the stage, pulse propagation along a waveguide with quadratic and quartic dispersion is governed by the Generalized Nonlinear Schr{\"o}dinger Equation (GNLSE) 
       \begin{equation}
       \frac{\partial A}{\partial z}=i \gamma |A|^{2} A - i \frac{\beta_{2}}{2}\frac{\partial ^{2} A}{\partial t^{2}}+i\frac{\beta_{4}}{24}\frac{\partial ^{4} A}{\partial t^{4}}
       \label{eq:gnlse}
       \end{equation}
for the complex pulse envelope $A(z,t)$; here $\beta_{2}$ and $\beta_{4}$ are the respective dispersion coefficients, and $\gamma$ represents the strength of the Kerr nonlinearity, which is cubic.

Theoretical and numerical work on the partial differential equation (PDE)~\eqref{eq:gnlse} has shown the persistence of a single-hump (pulse) soliton \cite{GDK}, and the existence of infinitely many solitons with different symmetry properties and numbers of humps  \cite{PhysRevA.103.063514, PARKER2021132890} for $\beta_4 < 0$ and a range of positive and negative values of $\beta_2$. These results are obtained by considering a traveling-wave ansatz, which transforms the PDE into a Hamiltonian and reversible system of ordinary differential equations (ODEs) \cite{devaney1977blue, devaney1976, champneys1998homoclinic}. For the GNLSE ~\eqref{eq:gnlse} with $A(z,t)=u(t)e^{i\mu z}$, one thus obtains the system of ODEs \cite{PhysRevA.103.063514}
\begin{equation}
	\cfrac{d\mathbf{u}}{dt}=\left( {\begin{array}{cc}u_{2} \\u_{3}\\u_{4}\\ \cfrac{24}{\beta_{4}}\left(\cfrac{\beta_{2}}{2}u_{3}+\mu u_{1}-\gamma u_{1}^3  \right) \\\end{array} } \right),
	\label{eq:ODEsystem}
\end{equation}
where $\mathbf{u}=(u_1,u_2,u_3,u_4)=\left(u,\frac{du}{dt},\frac{d^{2}u}{dt^{2}},\frac{d^{3}u}{dt^{3}}\right)$. System~\eqref{eq:ODEsystem} is Hamiltonian (thus, also volume-preserving) with the energy function \cite{PhysRevA.103.063514, PARKER2021132890} 
\begin{equation}
     H(\mathbf{u})=u_{2}u_{4}-\frac{1}{2}u_{3}^{2}-\left(\frac{6\beta_{2}u_{2}^{2}-6\gamma u_{1}^{4}+12\mu u_{1}^{2}}{\beta_{4}}\right) \nonumber
\end{equation}
as conserved quantity. Moreover, the system of ODEs~\eqref{eq:ODEsystem} has two reversibilities given by the transformations
\begin{eqnarray}
	R_{1}: & (u_{1},u_{2},u_{3},u_{4}) & \rightarrow (u_{1},-u_{2},u_{3},-u_{4}),
\label{eq:R1symm} \\
	R_{2}: & (u_{1},u_{2},u_{3},u_{4}) & \rightarrow (-u_{1},u_{2},-u_{3},u_{4})
\label{eq:R2symm}
\end{eqnarray}
with corresponding reversibility sections 
\begin{eqnarray}
	\Sigma_{1}& = & \{\mathbf{u}\in \mathbb{R}^{4} : u_{2}=u_{4}=0  \} \quad \mathrm{and} 
\label{eq:Sigma1} \\
	\Sigma_{2} & = & \{\mathbf{u}\in \mathbb{R}^{4} : u_{1}=u_{3}=0  \}
\label{eq:Sigma}
\end{eqnarray}
that are (pointwise) invariant under $R_1$ and $R_2$, respectively \cite{devaney1977blue, champneys1998homoclinic,articleParra,champneys1993hunting}. 
Hence, if $\mathbf{u}(t)$ is a trajectory or solution of system \eqref{eq:ODEsystem} then both $R_{1}(\mathbf{u}(-t))$ and $R_{2}(\mathbf{u}(-t))$ are also solutions \cite{champneys1998homoclinic,articleParra,champneys1993hunting}. We say that $\mathbf{u}(t)$ is \emph{symmetric} if it is invariant as a set under $R_1$ and/or $R_2$. Furthermore,  we make the following further distinction of its symmetry properties: we say that $\mathbf{u}(t)$ is 
\begin{itemize}
	\item \emph{$R_1$-symmetric} when it is invariant under only $R_1$; here, $\mathbf{u}$ intersects $\Sigma_1$ transversally; \\[-6mm]
	\item \emph{$R_2$-symmetric} when it is invariant under only $R_2$; here, $\mathbf{u}$ intersects $\Sigma_2$ transversally. \\[-6mm]
	\item \emph{$R^{*}$-symmetric} when it is invariant under $R_1$ as well as $R_2$; here, $\mathbf{u}$ intersects both $\Sigma_1$ and $\Sigma_2$ transversally. \\[-6mm]
	\item \emph{non-symmetric} when it is not invariant (under either $R_1$ or $R_2$).
\end{itemize}
Furthermore, system \eqref{eq:ODEsystem} is equivariant under the state-space symmetry transformation
\begin{equation}
S: (u_{1},u_{2},u_{3},u_{4}) \rightarrow (-u_{1},-u_{2},-u_{3},-u_{4}), \nonumber
\end{equation}
which is a point reflection in the origin $\mathbf{0} = (0,0,0,0)$ and given by $S=R_{1} \circ  R_{2}=R_{2} \circ R_{1}$. Notice that invariance of a solution $\mathbf{u}(t)$ under $S$ does not necessarily imply $R^*$-symmetry; however, any $R^{*}$-symmetric solution is necessarily invariant under $S$. 

Solitons of the GNLSE~\eqref{eq:gnlse} are given by the $u_1$-component of homoclinic orbits to the origin $\mathbf{0}=(0,0,0,0)$ of system~\eqref{eq:ODEsystem}, as a result of the traveling-wave ansatz \cite{PhysRevA.103.063514, PARKER2021132890}. Note that $\mathbf{0}$ is an equilibrium in the zero-energy level $H \equiv 0$ and, moreover, $\mathbf{0} \in \Sigma_{1} \cap \Sigma_{2}$. Hence, when $\mathbf{0}$ is a saddle, homoclinic orbits to it (approaching $\mathbf{0}$ in both forward and backward time) may be $R_1$-symmetric, $R_2$-symmetric, $R^{*}$-symmetric or non-symmetric \cite{PhysRevA.103.063514}. In this Hamiltonian setting, homoclinic orbits lie in the same energy level as the equilibrium, and they are the limits of a family of periodic orbits (which are locally parametrized by the energy $H$). Moreover, any homoclinic orbit is structurally stable; that is, it persists under changes of parameters --- unless it undergoes a bifurcation). There has been extensive work on fourth-order, reversible, and Hamiltonian systems \cite{devaney1977blue, devaney1976, champneys1998homoclinic,HOMBURG2010379,champneys1993hunting,amick_toland_1992,champneys1993bifurcation} to understand the mechanism in which homoclinic solutions arise and disappear as parameters are varied. 

\begin{figure}[t!]
   \centering
   \includegraphics{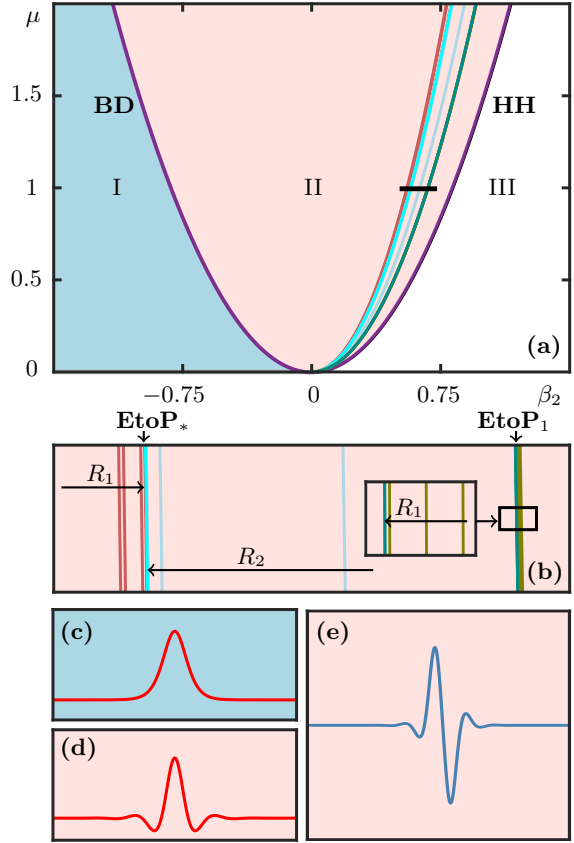} 
   \caption{Panel (a) shows the bifurcation diagram of \eqref{eq:ODEsystem} in the $(\beta_{2}, \mu)$-plane for $\beta_{4}=-1$ and $\gamma=1$ (a) with the curves \textbf{BD} and \textbf{HH} delimiting regions I (light blue), II (light green) and III (white); also shown in region II are fold curves of EtoP connections and associated solitons, whose ordering is illustrated in the enlargement panel~(b). Panels (c) and (d) show the primary $R_1$-symmetric homoclinic solution in region I for $(\beta_{2}, \mu)=(-1, 1)$ and for $(\beta_{2}, \mu)=(0.4, 1)$ in region II, respectively; panel~(e) shows the primary $R_2$-symmetric homoclinic solution in region II for $(\beta_{2}, \mu)=(0.4, 1)$.} 
       \label{fig:betamu}
\end{figure} 

System~\eqref{eq:ODEsystem} is known \cite{PhysRevA.103.063514,tam2018solitary,AKHMEDIEV1994540} to have a pair of primary $R_1$-symmetric homoclinic orbits to $\mathbf{0}$, which are each other's counterparts under the reversibility given by $R_2$. The eigenvalues of the equilibrium $\mathbf{0}$ become complex conjugate while the pair of homoclinic orbits exists, and this is known as a Belyakov-Devaney (\textbf{BD}) bifurcation \cite{devaney1977blue, devaney1976}. Moreover, the primary homoclinic orbit disappears in a Hamiltonian Hopf (\textbf{HH}) bifurcation when $\mathbf{0}$ ceases to be a saddle and becomes a center. Without loss of generality, we may consider the $(\beta_{2}, \mu)$-plane for $\beta_{4}=-1$ and $\gamma=1$ to find the relevant bifurcations, which all occur along semi-parabolas \cite{PhysRevA.103.063514}. Figure \ref{fig:betamu}(a) shows how the curves \textbf{BD} and \textbf{HH} together form a single (analytically known) parabola that divides the $(\beta_{2}, \mu)$-plane into region I to III \cite{PhysRevA.103.063514, GDK}. In regions I and II the equilibrium $\mathbf{0}$ is a real saddle and a saddle-focus, respectively. The primary $R_1$-symmetric homoclinic orbit exists in both these regions, and it is illustrated in panels~(c) and~(d) with temporal traces of $u_1$ that represent the corresponding soliton of the GNLSE~\eqref{eq:gnlse}. For simplicity we will refer to such $u_1$-traces representing solitons as \emph{homoclinic solutions} (as opposed to homoclinic orbits). Note that the soliton has non-oscillatory exponentially decaying tails in region I, and oscillating decaying tails in region II \cite{PhysRevA.103.063514,tam2018solitary,AKHMEDIEV1994540}. In region III, the equilibrium $\mathbf{0}$ is a center with purely imaginary eigenvalues and, hence, can no longer support homoclinic or other connecting orbits. 

The Belyakov-Devaney bifurcation \cite{devaney1977blue, devaney1976} implies that infinitely many homoclinic solutions of different symmetry types exist to the right of the curve \textbf{BD} in Fig.~\ref{fig:betamu}(a) \cite{PhysRevA.103.063514, champneys1993bifurcation}. Figure~\ref{fig:betamu}(e) shows the primary $R_2$-symmetric homoclinic solution, which is special because it emerges from \textbf{BD} and exists throughout region II; that is, it disappears only at the Hamiltonian Hopf bifurcation curve \textbf{HH}. All other homoclinic orbits arise from \textbf{BD} as infinite families of different symmetry types, and they are associated with heteroclinic connections between the equilibrium $\mathbf{0}$ and different periodic orbits, which we refer to as \emph{EtoP connections}. The EtoP connections also persist under parameter variation; they come in pairs that disappear (as $\beta_2$ is increased) along curves, again parabolas, of fold bifurcations in region II, some of which are shown in Fig.~\ref{fig:betamu}(a). As we will discuss in more detail in Sec.~\ref{sec:EtoP}, the associated homoclinic orbits of a given family also disappear in pairs at fold curves that are close to the fold curve of the corresponding EtoP connection \cite{PhysRevA.103.063514}; the enlargement in Fig.~\ref{fig:betamu}(b) illustrates the ordering of these fold curves. We refer to the overall bifurcation structure as \emph{BD-truncated homoclinic snaking}  (as opposed to regular homoclinic snaking \cite{woods1999heteroclinic,articleBruke}) and note that this phenomenon is also observed in a certain parameter regime of the Lugiato-Lefever equation \cite{articleParra}. 

In this paper, we establish the existence in the GNLSE~\eqref{eq:gnlse} of heteroclinic connections between saddle periodic orbits in the same energy surface, which we refer to as \emph{PtoP connections}. This type of connecting orbit also persists with parameters and plays an important role in the GNLSE for two main reasons. First of all, PtoP connections are organizing centers that `generate' associated families of homoclinic orbits to saddle periodic orbits. Each such homoclinic orbit corresponds to a soliton with non-decaying periodic tails. This type of solution to a `periodic background' is referred to as a \emph{generalized soliton} in the context of optics \cite{Alexander:22}. Much like EtoP connections organize solitons with decaying tails, PtoP connections are shown to give rise to infinite families of generalized solitons with different symmetry properties. 

Secondly, PtoP connections can be found in the zero-energy level. This allows us to combine them with EtoP connections to obtain heteroclinic cycles that, in turn, give rise to additional infinite families of solitons with decaying tails. These solitons are characterized by the corresponding homoclinic orbits `visiting' not just one but several periodic orbits: they follow an EtoP connection from $\mathbf{0}$ to a periodic orbit, perform a certain number of oscillations near it, then follow a PtoP connection to another periodic orbit, perform a certain number of oscillations near it, possible follow another PtoP connection, and so on, until they finally return to $\mathbf{0}$ via another EtoP connection. In contrast to the solitons considered before \cite{PhysRevA.103.063514}, which visit a single periodic orbit, we refer to these solitons organized by PtoP connections as \emph{multi-oscillation solitons}. As we will show, they are also created at \textbf{BD}, may have different symmetry properties, and disappear in fold bifurcations before \textbf{HH} is reached. Hence, multi-oscillation solitons are an integral part of the phenomenon of BD-truncated homoclinic snaking in system~\eqref{eq:ODEsystem}. 

Similarly, PtoP connections in the zero-energy surface arise in pairs from the Belyakov-Devaney bifurcation \textbf{BD} and disappear in fold bifurcations before \textbf{HH} is reached. Each such pair of PtoP connections gives rise to infinite families of homoclinic orbits of different symmetry properties to (each of) the two periodic orbits involved. In other words, we also find BD-truncated homoclinic snaking of homoclinic orbits to periodic orbits that are organized by corresponding PtoP connections. This phenomenon for homoclinic orbits to periodic orbits, which has not been reported previously in the literature, effectively generalizes BD-truncated homoclinic snaking of homoclinic orbits to equilibria \cite{PhysRevA.103.063514, articleParra}. On the other hand, PtoP connections are not confined to the zero-energy level, and they form surfaces in phase space. We compute and present such surfaces, and this provides a connection to theoretical results on the existence of surfaces of homoclinic solutions to periodic solutions in reversible systems \cite{homburg_lamb_2006}. 

Clearly, PtoP connections require and go hand-in-hand with periodic orbits that they connect. Periodic orbits also form surfaces in phase space that are locally parametrized by the Hamiltonian function of the ODE \cite{devaney1977blue, devaney1976}. A central role in the overall organization of connecting obits is played by the families of the periodic orbits that have as limits the pair of primary homoclinic orbits. Initially, for $\beta_2$ in region II and not too far from the curve \textbf{BD} in Fig.~\ref{fig:betamu}(a), there are three disjoint basic surfaces of periodic solutions. We compute these surfaces and show that they each intersect the zero-energy level infinitely many times. The overall picture of the soliton structure of the GNLSE~\eqref{eq:gnlse} that therefore emerges is that the pair of primary homoclinic orbits generates infinitely many saddle periodic orbits in the zero-energy level. Sufficiently close to \textbf{BD} in region II, each such periodic orbit generates a pair of EtoP connections of $\mathbf{0}$. Moreover, any two periodic orbits in the zero-energy level can be connected by a pair of PtoP connections, which generate infinite families of homoclinic orbits to either of the periodic orbits. Concatenating any pair of EtoP connections with any number of PtoP connections creates infinitely many heteroclinic cycles from and to $\mathbf{0}$, each of which generates an entire menagerie of additional and arbitrarily complicated homoclinic orbits to the origin --- and, hence novel types of solitons of the GNLSE. In turn, each homoclinic orbit is the limit of families of periodic orbits, infinitely many of which lie in the zero-energy level so that they also generate pairs of EtoP and PtoP solutions and so on.

The work presented here makes heavy use of state-of-the-art computational methods that allow us to find different periodic, homoclinic and heteroclinic orbits, to continue them in parameters and to detect their bifurcations. In particular, we formulate connecting orbits between saddle objects as solutions of suitably defined two-point boundary value problems (2PBVPs) \cite{krauskopf2007numerical,PhysRevA.103.063514}. In an approach called Lin's method \cite{krauskopf2008lin}, this involves finding intersections of computed parts of stable and unstable manifolds of equilibria and periodic orbits in a three-dimensional section. Since system~\eqref{eq:ODEsystem} is Hamiltonian, it is necessary to ensure that solutions to the respective 2PBVPs are isolated, which is achieved by introducing the gradient of the conserved quantity $H$ multiplied with a perturbation parameter \cite{galan2014continuation,PhysRevA.103.063514} to system~\eqref{eq:ODEsystem}. All these computations are performed with the continuation package \textsc{Auto-07p} \cite{doedel2007auto} and its extension \textsc{HomCont} \cite{champneys1996numerical}.  

The paper is structured as follows. In Sec.~\ref{sec:basic}, we introduce the three basic surfaces $\mathcal{S}_{1}^{\pm}$ and $\mathcal{S}_{*}$ of periodic orbits that accumulate on the primary homoclinic orbit and its $R_2$-counterpart. In Sec.~\ref{sec:EtoP}, we then briefly summarise previous results \cite{PhysRevA.103.063514} regarding how EtoP connections to specific periodic orbits in these surfaces organize associated families of homoclinic orbits to $\mathbf{0}$; here, we also show a one-parameter bifurcation diagram that clarifies over which $\beta_2$-ranges they exist. This sets the stage for the discussion in Sec.~\ref{sec:PtoPgeneral} of PtoP connections and how they generate generalized solitons. We first consider PtoP connections in the zero-energy, where they exist, and how they organize corresponding families of homoclinic orbits with different symmetry properties. We also calculate and present the surfaces of these PtoP connections, which are not restricted to have zero energy. We then provide in Sec.~\ref{sec:PtoPmulti} a schematic diagram that shows how EtoP and PtoP connections can be concatenated into heteroclinic cycles. Specifically, we show how families of homoclinic orbits to $\mathbf{0}$ with different symmetry properties are generated by combining different EtoP and PtoP connections. These additional types of homoclinic orbits are indeed multi-oscillation solitons of the GNLSE, and one-parameter bifurcation diagrams show their $\beta_2$-ranges of existence. We conclude in Sec.~\ref{sec:conclusions} with a brief summary and an outlook on open questions and future research. 

%
\section{The surfaces of basic periodic orbits}
\label{sec:basic}

\begin{figure*}[t!]
   \centering
   \includegraphics{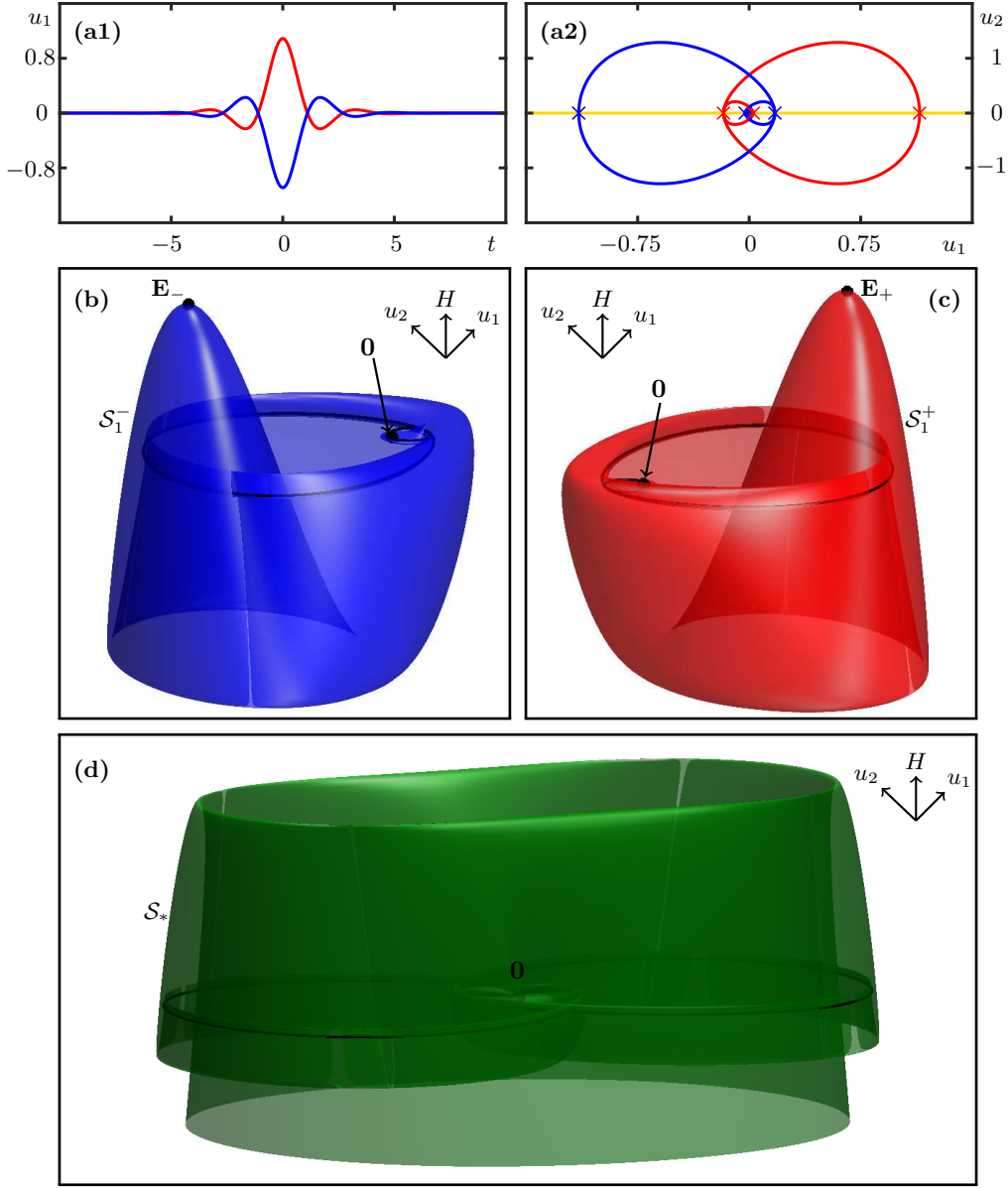} 
   \caption{The primary $R_1$-symmetric homoclinic orbit to $\mathbf{0}$ (red) and its $R_2$-counterpart (blue) of system \eqref{eq:ODEsystem} for $\beta_2 = 0.4$, shown in panel (a1) as temporal profiles of the $u_1$-component and in projection onto the $(u_1,u_2)$-plane, respectively. They are shown in $(u_1,u_2,H)$-space with the respective surfaces of periodic orbits, namely, $\mathcal{S}_{1}^{-}$ (blue) with equilibrium $\mathbf{E^{-}}$ in panel (b),  $\mathcal{S}_{1}^{+}$ (red) with equilibrium $\mathbf{E^{+}}$ in panel (c), and $\mathcal{S}_{*}$ (green) in $(u_1,u_2,H)$-space in panel (d). Throughout, the other parameters are set to $(\beta_4,\gamma,\mu)=(-1,1,1)$.}
     \label{fig:surfaces}
\end{figure*} 

The primary homoclinic orbit from Fig.~\ref{fig:betamu} is $R_1$-symmetric but not $R_2$-symmetric, meaning that it has an $R_2$-counterpart. As is illustrated in Fig.~\ref{fig:surfaces}, for $\beta_2 = 0.4$, this pair of primary homoclinic solutions gives rise to three basic surfaces; here and throughout, we fix, without loss of generality, $\beta_4=-1$ and $\gamma=1$, and set $\mu = 1$ since all bifurcation curves in the $(\beta_2,\mu)$-plane are parabolas; note here that Eq.~\eqref{eq:gnlse} can be rescaled appropriately for different signs of $\beta_2$ and $\beta_4$, as was shown previously \cite{PhysRevA.103.063514}. Figure~\ref{fig:surfaces}(a1) shows the pair of primary homoclinic orbits in terms of their $u_1$-profile, while panel~(a2) shows them in the $(u_1,u_2)$-plane. They each come with a surface of periodic orbits that accumulates on the respective homoclinic orbit in the zero-energy level. These two surfaces $\mathcal{S}_{1}^{-}$ and $\mathcal{S}_{1}^{+}$ are shown in projection onto $(u_1,u_2,H)$-space in panels (b) and (c), respectively. The parametrization by the energy $H$ is explicit, and this allows us to easily identify the periodic orbits in the zero-energy level, which contains the origin $\mathbf{0}$ and the pair of primary homoclinic orbits, which are also shown in Fig.~\ref{fig:surfaces}(b) and (c). Note that $\mathcal{S}_{1}^{-}$ and $\mathcal{S}_{1}^{+}$ are both $R_1$-symmetric and $R_2$-counterparts of one another \cite{PhysRevA.103.063514}. Moreover, they each extend over only a bounded range of $H$, where the largest value of $H$ occurs at the equilibria $\mathbf{E^{-}}$ and $\mathbf{E^{+}}$. These are centers at $\mathbf{E_{\pm}}=\left(\pm \sqrt \frac{\mu}{\gamma},0,0,0\right)$ (so only exists if $\mu \gamma>0$) and each others $R_2$-counterparts with $H(\mathbf{E_{\pm}})=-6\mu^2/(\gamma\beta_{4})$. 
The union of the pair of primary homoclinic orbits is the limit of a third, $R_*$-symmetric surface $\mathcal{S}_{*}$, which is also shown in $(u_1,u_2,H)$-space in Fig.~\ref{fig:surfaces}(d) together with the pair of primary homoclinic orbits. This surface has a maximal value of $H$ but extends to arbitrarily large negative values of $H$; that is, it is not bounded from below \cite{PhysRevA.103.063514}. 

\begin{figure*}[t!]
   \centering
   \includegraphics{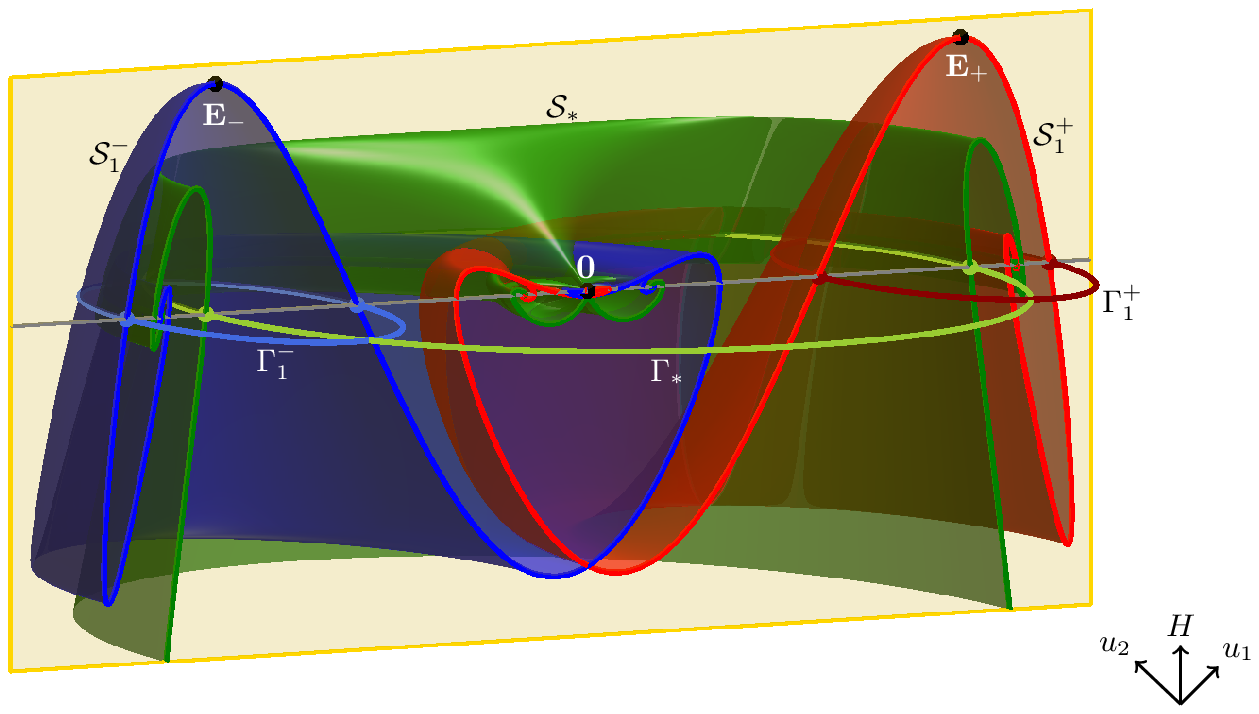} 
   \caption{The surfaces $\mathcal{S}_{*}$, $\mathcal{S}_{1}^{-}$ and $\mathcal{S}_{1}^{+}$ from Fig.~\ref{fig:surfaces}(b)--(d), shown in a cutaway view in $(u_1,u_2,H)$-space up to the plane $\Sigma$ where $u_2=0$; also shown are the equilibria $\mathbf{0}$ and $\mathbf{E_{\pm}}$ (black dots), the zero-energy level (grey line) in $\Sigma$, and the primary periodic orbits $\Gamma_{1}^{-}$ (blue curve), $\Gamma_{1}^{+}$ (red curve) and $\Gamma_{*}$ (green curve) in the zero-energy level.} 
\label{fig:cutaway}
\end{figure*} 

We refer to the the surfaces $\mathcal{S}_{1}^{\pm}$ and $\mathcal{S}_{*}$ as the surfaces of \emph{basic periodic orbits}, and they are shown together in
Fig.~\ref{fig:cutaway} in projections onto $(u_1,u_2,H)$-space, namely in a cutaway view that shows their parts for $u_2 \leq 0$ up to the section $\Sigma = \{ (u_1, 0, u_3, u_4) \}$ defined by $u_2=0$. Notice that the respective other half, for positive $u_2$ on the other side of $\Sigma$, is obtained from that shown by applying the transformation $R_1$. In Fig.~\ref{fig:cutaway}, this corresponds to a reflection in the $\Sigma$-plane; compare with Fig.~\ref{fig:surfaces}(b)--(d). 

This cutaway representation allows us to compute and show the intersection curves of the surfaces $\mathcal{S}_1^{\pm}$ and $\mathcal{S}_{*}$ with the plane $\Sigma$, which is very useful for understanding the geometry of these surfaces. The gray line in $\Sigma$ indicates the zero-energy level with $u_2=0$, which contains the basic homoclinic orbits. As Fig.~\ref{fig:cutaway} illustrates, the basic surfaces $\mathcal{S}_1^{\pm}$ and $\mathcal{S}_{*}$ each spiral as they limit on the (union of) the two basic homoclinic orbits, which means that they intersect the zero-energy level infinitely many times. To be more specific, starting from the equilibria $\mathbf{E_{\pm}}$ and decreasing the energy $H$, one encounters in the zero-energy level the $R_1$-symmetric and $R_2$-related pair of periodic orbits $\Gamma_{1}^{-}$ on $\mathcal{S}_1^{-}$ and $\Gamma_{1}^{+}$ on $\mathcal{S}_1^{+}$. Similarly, starting from large negative $H$ and increasing the energy, one encounters in the zero-energy level the $R_*$-symmetric periodic orbit $\Gamma_{*}$ on $\mathcal{S}_{*}$. The three periodic orbits $\Gamma_{1}^{\pm}$ and $\Gamma_{*}$ are shown in full in Fig.~\ref{fig:cutaway}. As can be seen, they are the `first' periodic orbits one encounters in the zero-energy level as the respective surface accumulates on the pair of homoclinic orbits. For this reason, we refer to $\Gamma_{1}^{\pm}$ and $\Gamma_{*}$ as the \emph{primary periodic orbits}; note, in particular, that each of them is a single-loop orbit with exactly two intersection points with the section $\Sigma$. 

In this work, we demonstrate with the specific examples of the primary periodic orbits $\Gamma_{1}^{\pm}$ and $\Gamma_{*}$ for $\beta_2 = 0.4$ how connecting orbits between them give rise to PtoP connection, which in turn generate generalized and multi-oscillation solitons. Indeed, the same is true for the further periodic orbits of $\mathcal{S}_1^{\pm}$ and $\mathcal{S}_{*}$ with $H=0$, of which there are countably infinitely many. In contrast to the primary periodic orbits, further periodic orbits of $\mathcal{S}_1^{\pm}$ and $\mathcal{S}_{*}$ in the zero-energy level are multi-loop orbits with an increasing number of additional intersection points with the (three-dimensional) section $\Sigma$. How they arise from the associated structure of $\mathcal{S}_1^{\pm}$ and $\mathcal{S}_{*}$, and how these surfaces depend on the parameter $\beta_2$ are interesting subjects beyond the scope of this paper that will be discussed elsewhere \cite{bgbk_periodic}.

\section{E{\scriptsize to}P cycles and their associated solitons}
\label{sec:EtoP}

As a starting point for our introduction of PtoP orbits, we briefly recall and discuss here what families of solitons of different symmetry types are generated in region II by EtoP connections from $\mathbf{0}$ to the primary periodic orbits $\Gamma_{1}^{\pm}$ and $\Gamma_{*}$ \cite{PhysRevA.103.063514}, again for $\beta_2 = 0.4$ (and $\mu=1$). The dependence of the EtoP connections and associated solitons on $\beta_2$ is subsequently presented as a one-parameter bifurcation diagram, which introduces the different fold curves already shown in Fig.~\ref{fig:betamu}.

\begin{figure*}[t!]
   \centering
   \includegraphics{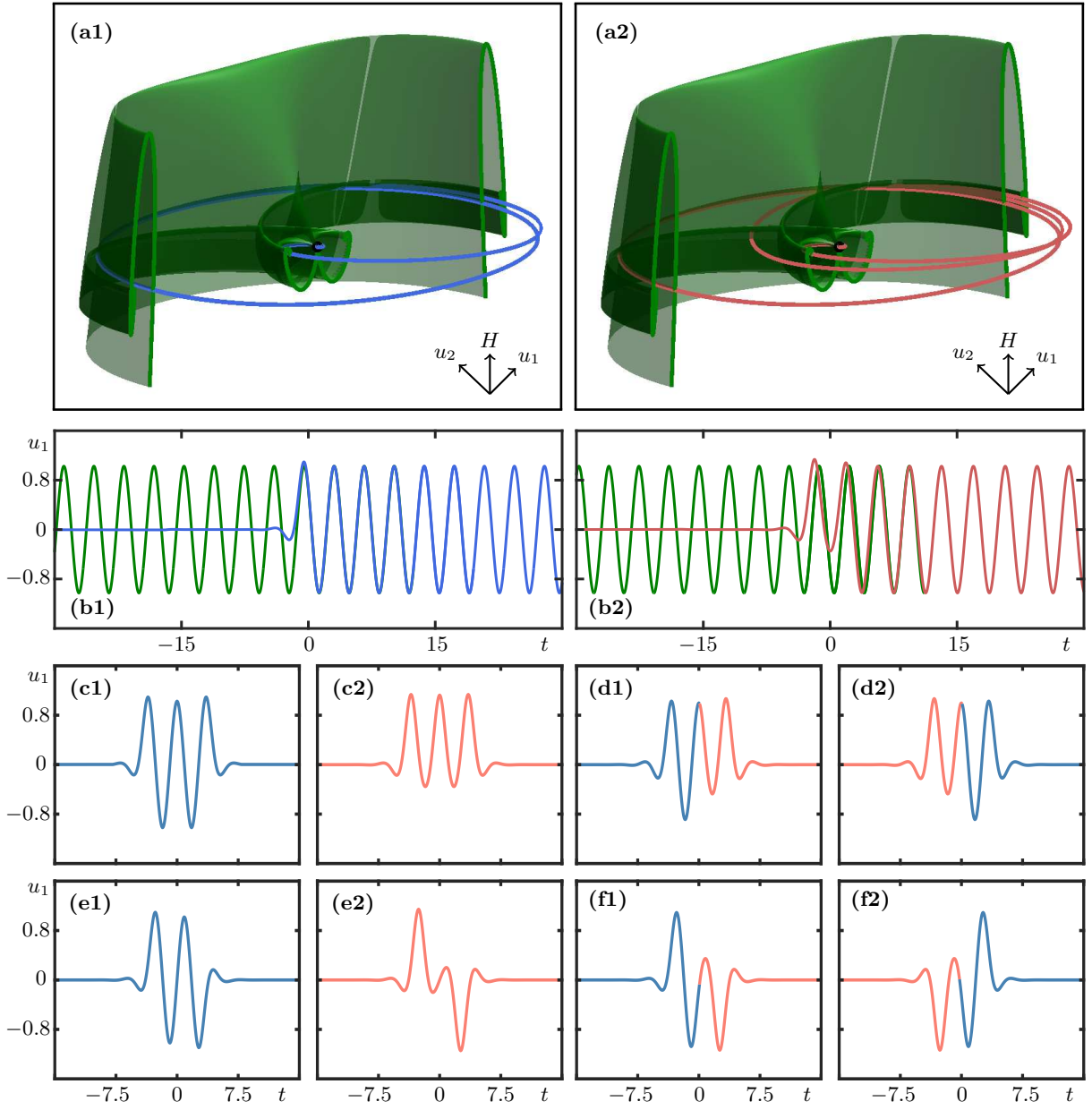} 
   \caption{Homoclinic solutions associated with the basic $R_{*}$-symmetric periodic orbit $\Gamma_{*}$ for $\beta_2=0.4$. Panels (a1) and (b1) show two distinct EtoP connections to $\Gamma_{*}$ in $(u_1, u_2, H)$-space with a cutaway view of 
the surface $\mathcal{S}_{*}$; and panels~(a2) and (b2) show their $u_1$-traces. Panels~(c1), (c2) and (d1), (d2) show selected $R_{1}$-symmetric and $R_1$-symmetry-broken homoclinic solutions; and panels~(e1), (e2) and (f1),(f2) show selected $R_{2}$-symmetric and $R_{2}$-symmetry-broken homoclinic solutions. Different coloring indicates the parts of the homoclinic orbits that are close to the respective EtoP connection (or its image under $R_1$ or $R_2$).}
       \label{fig:solG*}
\end{figure*} 

\begin{figure*}[t!]
   \centering
   \includegraphics{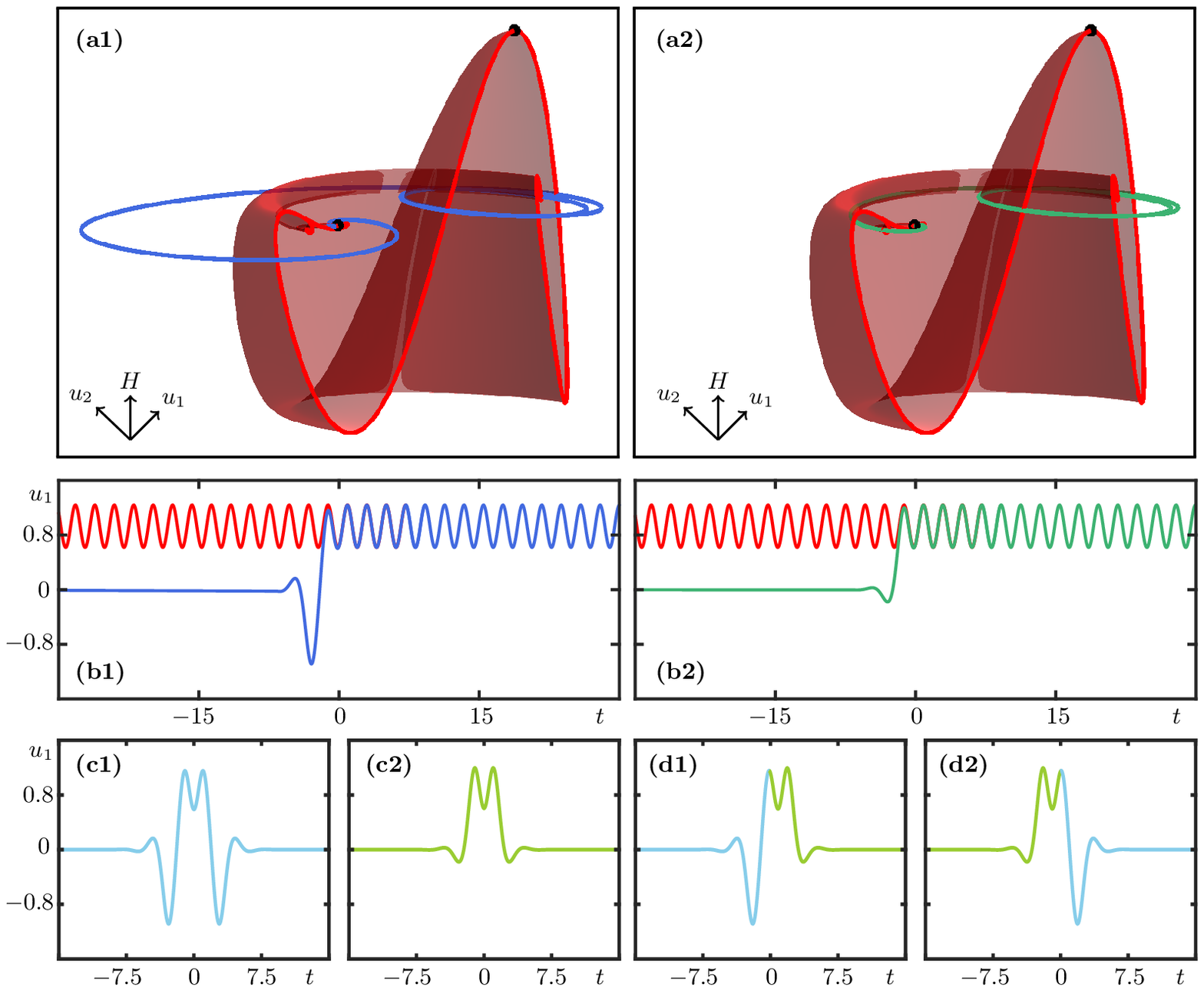} 
   \caption{Homoclinic solutions associated with the primary $R_{1}$-symmetric periodic orbit $\Gamma_{1}^{+}$ for $\beta_2=0.4$. Panels (a1) and (b1) show the two distinct EtoP connections to $\Gamma_{1}^{+}$ in $(u_1, u_2, H)$-space with a cutaway view of the surface $\mathcal{S}_{1}^{+}$; and panels~(a2) and (b2) show their $u_1$-traces. Panels~(c1), (c2) and (d1),(d2) show selected $R_{1}$-symmetric and $R_1$-symmetry-broken homoclinic solutions, respectively. Different coloring indicates the parts of the homoclinic orbits that are close to the respective EtoP connection (or its image under $R_1$).}
       \label{fig:solG1pm}
\end{figure*} 

Figures~\ref{fig:solG*} and~\ref{fig:solG1pm} show how the EtoP connections to $\Gamma_{*}$ and to $\Gamma_{1}^{\pm}$, together with their corresponding symmetric counterparts, generate heteroclinic cycles that organize homoclinic orbits to $\mathbf{0}$ under mild conditions \cite{palis2012geometric,articleCham,PhysRevA.103.063514, PARKER2021132890}. These homoclinic orbits closely follow an EtoP connection and make a number of loops around the respective periodic orbit before converging back to $\mathbf{0}$ along a return EtoP connection. Homoclinic orbits with any number of loops can be found, and they form infinite families with increasing numbers of `humps' of the corresponding solitons. Since system \eqref{eq:ODEsystem} possesses two reversible symmetries, families of homoclinic orbits with different symmetry properties can be obtained through this mechanism. 

Figure~\ref{fig:solG*} shows the EtoP connections to the primary periodic orbit $\Gamma_{*}$ and associated homoclinic solutions. Panels~(a) and~(b) show two EtoP connections from $\mathbf{0}$ to $\Gamma_{*}$ that exist simultaneously for $\beta_2=0.4$. Here, panels~(a1) and~(b1) show the EtoP connections in $(u_1, u_2, H)$-space together with a cutaway view of 
the surface $\mathcal{S}_{*}$; and panels~(a2) and (b2) show their $u_1$-traces, which illustrates that the EtoP connections converge backward in time to $\mathbf{0}$ and forward in time to $\Gamma_{1}^{+}$. Note that these two EtoP connections are distinct in the sense that they are not related by any of the symmetries of \eqref{eq:ODEsystem}. However, since $\Gamma_{*}$ is $R^{*}$-symmetric, the $R_1$-counterparts as well as the $R_2$-counterparts of these two EtoP connections exist as well. Note, in particular, that the $R_2$-counterparts provide return connections from $\Gamma_{*}$ to $\mathbf{0}$. Hence, the EtoP connections in Fig.~\ref{fig:solG*}(a) and (b) together with their corresponding $R_1$- and $R_2$-counterparts generate $R_1$-symmetric and $R_1$-symmetry-broken, as well as $R_2$-symmetric and $R_2$-symmetry-broken homoclinic orbits, respectively \cite{PhysRevA.103.063514}. Some representative examples are shown in Fig.~\ref{fig:solG*}(c)--(f) in terms of their $u_1$-traces, where we distinguished by color the two parts that follow a particular EtoP connection (or its symmetric counterpart). More specifically, panels~(c1) and~(c2) show solutions with $R_1$-symmetry that are generated from each of the two EtoP connections and their $R_1$-counterparts; here, each EtoP connection is effectively followed to its second maximum. Similarly, an EtoP connection and then the $R_1$-counterpart of the other combine to the $R_1$-symmetry-broken homoclinic solutions shown in panels~(d1) and~(d2). In complete analogy, panels~(e1) and~(e2) show homoclinic orbits with $R_2$-symmetry that are generated from each of the two EtoP connections and their $R_2$-counterparts. Combining an EtoP connection with the $R_2$-counterpart of the other gives the $R_2$-symmetry-broken homoclinic orbits shown in panels~(f1) and~(f2). We remark that the images of these solitons under the reversibility transformations given by $R_1$ and $R_2$ also exist; in the representation of the $u_1$-trace in Fig.~\ref{fig:solG*}(c)--(f), these transformations are geometrically: reflection in the $u_1$ axis, and rotation by $\pi$ around the origin of the $(t,u_2)$-plane, respectively.

In the same style, Fig.~\ref{fig:solG1pm} shows the EtoP connections to $\Gamma_{1}^{+}$ and associated homoclinic solutions that exist simultaneously. The difference is now that $\Gamma_{1}^{+}$ is not $R_2$-symmetric, and this means that the two distinct EtoP connections to $\Gamma_{1}^{+}$, which are shown in panels~(a) and~(b) with a cutaway view of $\mathcal{S}_{1}^{+}$, can only be combined with their $R_1$-counterparts to generate heteroclinic cycles between $\mathbf{0}$ and $\Gamma_{1}^{+}$. (Of course, their images under the reversibility $R_2$ also exist, but they are heteroclinic cycles between $\mathbf{0}$ and $\Gamma_{1}^{-}$.) Panels~(c1) and~(c2) of Fig.~\ref{fig:solG1pm} show examples of $R_1$-symmetric solitons that follow the respective EtoP connection to the first minimum near $\Gamma_{1}^{+}$, and then follow the $R_1$-counterpart back to $\mathbf{0}$. In contrast, the solitons shown in panels~(d1) and~(d2) are composed by each of the EtoP connections in combination with the $R_1$-counterpart or the other EtoP connection, and this yields $R_1$-symmetry-broken solitons.

\begin{figure}[t!]
   \centering
   \includegraphics{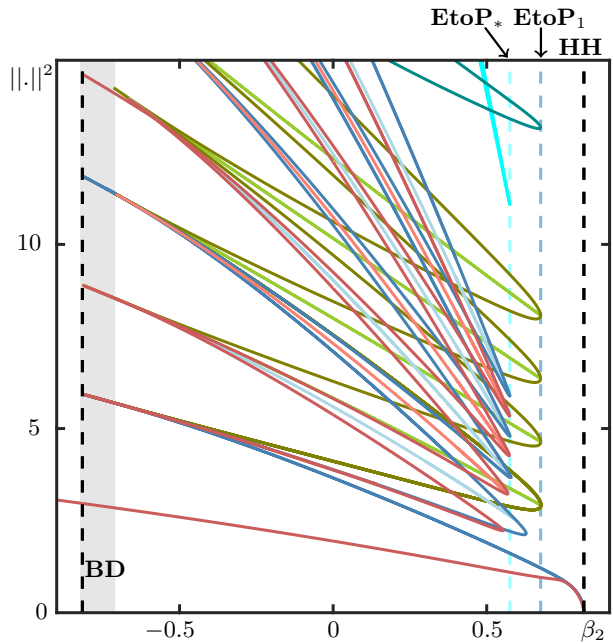} 
   \caption{One-parameter bifurcation diagrams in $\beta_2$ of the families of homoclinic orbits associated with branches of the EtoP connections to $\Gamma_{*}$ (cyan) and to $\Gamma_{1}^{+}$ (maroon). Shown are branches of $R_1$-symmetric (dark red), $R_1$-symmetry-broken (light red), $R_2$-symmetric (dark blue) and $R_1$-symmetry-broken (light blue) solitons with oscillations near $\Gamma_{*}$; and $R_1$-symmetric (dark green) and $R_1$-symmetry-broken (light green) solitons with oscillations near $\Gamma_{1}^{+}$; compare with Figs.~\ref{fig:solG*} and~\ref{fig:solG1pm}. The vertical dashed lines indicate the bifurcations \textbf{BD} and \textbf{HH}, and the fold points \textbf{EtoP$_*$} and \textbf{EtoP$_1$} of the curves of EtoP connections to $\Gamma_{*}$ and $\Gamma_{1}^{+}$, respectively.}
       \label{fig:bifEtoP}
\end{figure} 

Because system \eqref{eq:ODEsystem} is reversible and Hamiltonian, all connecting orbits, the EtoP connections and associated homoclinic orbits, persist in open regions of parameter space \cite{champneys1998homoclinic,ELVIN2010537}. Hence, once found, each such object can be continued as a solution branch in a chosen parameter, which we take to be the strength $\beta_2$ of the quadratic dispersion. In this way, one can compute a one-parameter bifurcation diagram that summarizes over what ranges of $\beta_2$ which types of connections and associated solitons can be found \cite{PhysRevA.103.063514}. Figure~\ref{fig:bifEtoP} shows this kind of bifurcation diagram for the EtoP connections and different symmetric and symmetry-broken homoclinic orbits from Figs.~\ref{fig:solG*} and~\ref{fig:solG1pm}. Here the shown $\beta_2$-range includes the entire region II bounded by the bifurcations \textbf{BD} and \textbf{HH}, which are indicated by vertical lines and lie at $\beta_2 \approx \pm 0.8164$ (for $\mu = 1$). The different branches are represented by their $L_2$-norm $|| . ||^2$, which is computed as an integral over the respective computed (part of the) connecting orbit \cite{doedel2007auto}. The $L_2$-norm is a good choice to represent and distinguish branches of solitons because it increases with the number of maxima of the corresponding homoclinic orbits. 

As Fig.~\ref{fig:bifEtoP} shows, the two EtoP connections to $\Gamma_{*}$ from Figs.~\ref{fig:solG*}(a) and~(b) lie on branches that emerge from \textbf{BD} and meet at a fold point at $\beta_2 \approx 0.5753$ to form a single smooth curve. The location of this fold, which we refer to as \textbf{EtoP$_*$}, is highlighted in Fig.~\ref{fig:bifEtoP} by a vertical dashed line. The primary $R_1$-symmetric homoclinic orbit exists throughout regions I and II, forming a single branch that crosses \textbf{BD} and ends at \textbf{HH}. Note from Fig.~\ref{fig:betamu}(d) that, in region II, it can be interpreted as being part of the family of $R_1$-symmetric homoclinic orbits that are close to an EtoP cycle between $\mathbf{0}$ and $\Gamma_{*}$. The other homoclinic orbits of this family, with more and more maxima as in Fig.~\ref{fig:solG*}(c), lie on branches that start at \textbf{BD} and meet in pairs at fold points to form smooth curves. In fact, these folds are points of symmetry breaking: they are also fold points where corresponding branches of $R_1$-symmetry-broken homoclinic orbits, as those in Fig.~\ref{fig:solG*}(d), meet to form single curves from and back to \textbf{BD}. Four curves of each family are shown in Fig.~\ref{fig:bifEtoP}; note that two simultaneously existing symmetry-broken solutions have the same $L_2$-norm, which is why their two branches cannot be distinguished in Fig.~\ref{fig:bifEtoP}. The situation for the families of $R_2$-symmetric and $R_2$-symmetry-broken homoclinic orbits is very similar. The primary $R_2$-symmetric homoclinic orbit from Fig.~\ref{fig:betamu}(e) is created at \textbf{BD} and ends at \textbf{HH}, and it can be interpreted as being part of the family of $R_2$-symmetric homoclinic orbits that follow an EtoP cycle between $\mathbf{0}$ and $\Gamma_{*}$. As Fig.~\ref{fig:bifEtoP} shows, all other $R_2$-symmetric homoclinic orbits, such as those in Fig.~\ref{fig:solG*}(e), come in pairs that form single curves, which meet the curves of the corresponding pairs of $R_2$-symmetry-broken homoclinic orbits, as those in Fig.~\ref{fig:solG*}(e), at their respective fold points. Again, four curves of these families are shown in Fig.~\ref{fig:bifEtoP}, and they illustrate that the folds of the families of $R_1$-symmetric and of the $R_2$-symmetric homoclinic orbits converge to the fold \textbf{EtoP$_*$}, from the left and from the right, respectively; compare with Fig.~\ref{fig:betamu}(b). 

Similarly, the two EtoP connections to $\Gamma_{1}^{+}$ from Figs.~\ref{fig:solG1pm}(a) and~(b) (and also those to $\Gamma_{1}^{-}$ by $R_2$-symmetry) lie on a single branch that emerges from and returns to  \textbf{BD}, with a fold point that lies at $\beta_2 \approx 0.6756$. We refer to this fold as \textbf{EtoP$_1$}, and it is also highlighted by a vertical dashed line in Fig.~\ref{fig:bifEtoP}. For this type of $R_1$-symmetric EtoP connection, there only exist $R_1$-symmetric and $R_1$-symmetry-broken homoclinic orbits that come in pairs, as those in Fig.~\ref{fig:solG1pm}(c) and~(d). Their branches, of which four are shown in Fig.~\ref{fig:bifEtoP}, meet in the same way at joint fold points. These fold points are very close to and converge to the fold \textbf{EtoP$_1$} from the right; see Fig.~\ref{fig:betamu}(b).

\section{P{\scriptsize to}P connections and generalized solitons}
\label{sec:PtoPgeneral}

We now show that one can find pairs of heteroclinic connections between two given saddle periodic orbits in the same energy level. More specifically, we compute and present such PtoP connections between the saddle periodic orbits $\Gamma_{*}$ and $\Gamma_{1}^{+}$, and show that they emerge from the Belyakov-Devaney bifurcation \textbf{BD} and exist over a considerable $\beta_2$-range in region II.

\begin{figure}[t!]
   \centering
   \includegraphics{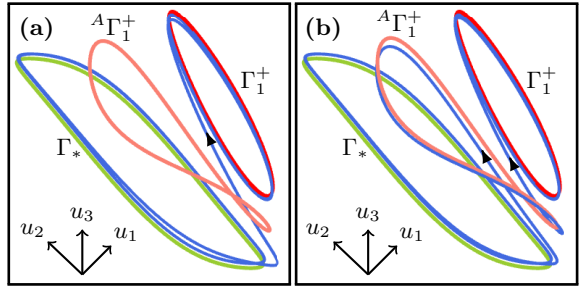} 
   \caption{Pair of PtoP connections (blue curves) between $\Gamma_{*}$ (green curve) to $\Gamma^{+}_1$ (red curve) of system \eqref{eq:ODEsystem} for $\beta_2= 0.4$. Panels~(a) and~(b) show each of the PtoP connections in $(u_1,u_1,u_3)$-space, where the arrows indicate the direction of time; also shown is the additional saddle periodic orbit $^{A}\Gamma^{+}_1$ (salmon curve) that also lies on the surface $\mathcal{S}_1^{+}$.}  
      \label{fig:PtoPphase}
\end{figure} 

Figure~\ref{fig:PtoPphase} shows in projection onto $(u_1,u_1,u_3)$-space a pair of PtoP connections between $\Gamma_{*}$ and $\Gamma_{1}^{+}$ for $\beta_2= 0.4$. Each of these two PtoP connections is clearly seen to converge backward in time to $\Gamma_{*}$ and forward in time to $\Gamma^{+}_1$, and they are not related by symmetry. While the PtoP connection in Fig.~\ref{fig:PtoPphase}(a) transitions swiftly from $\Gamma_{*}$ to $\Gamma^{+}_1$, the one in panel~(b) has almost a complete loop near and around a third saddle periodic orbit $^{A}\Gamma^{+}_1$ that also lies on the surface $\mathcal{S}_1^{+}$. All objects shown in Fig.~\ref{fig:PtoPphase} lie in the zero-energy level.

As we discuss next, such pairs of PtoP connections give rise to families of homoclinic orbits, to either $\Gamma_{*}$ or $\Gamma_{1}^{\pm}$ in this case. In the GNLSE, they correspond to generalized solitons \cite{Alexander:22} with non-decaying periodic tails, which are not confined to the zero-energy level. In Sec.~\ref{sec:PtoPmulti}, we then show how PtoP connections in the zero-energy level can be combined with EtoP connections to form heteroclinic cycles from and back to $\mathbf{0}$ that generate families of multi-oscillation solitons.

\subsection{Homolinic orbits to $\Gamma_*$ }\label{sec:HomGamma*}

\begin{figure*}[t!]
   \centering
   \includegraphics{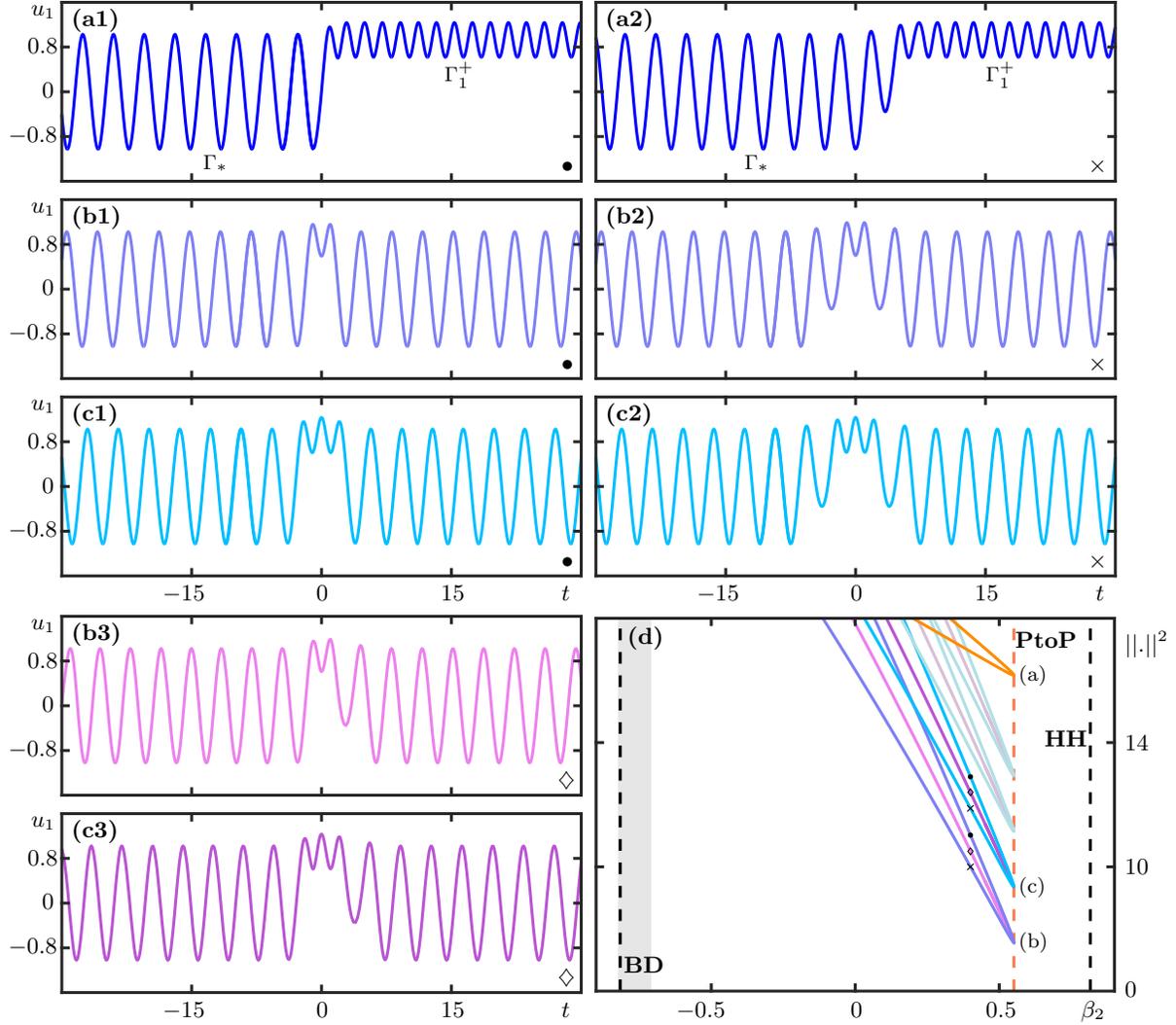} 
   \caption{Pair of PtoP connections from $\Gamma_{*}$ to $\Gamma^{+}_1$ and associated families of homoclinic solutions to $\Gamma_{*}$. Panels~(a1) and~(a2) show the two PtoP connections; panels (b1),(c1) and (b2),(c2) $R_1$-symmetric, and panels (b3),(c3) $R_1$-symmetry-broken homoclinic solutions to $\Gamma_{*}$. Panel (d) is the corresponding one-parameter bifurcation diagram in $\beta_2$ with the curve of PtoP connections (orange curve) and curves of homoclinic connections to $\Gamma_{*}$. The vertical dashed lines indicate \textbf{BD}, \textbf{HH} and the fold point \textbf{PtoP} of PtoP connections; the color of the bifurcation curves and the dots, crosses and diamonds indicate the locations of the solutions in panels (b1) and (c1), (b2) and (c2), and (b3) and (c3), respectively.} 
       \label{fig:HomGamma*}
\end{figure*} 

Figure~\ref{fig:HomGamma*} shows families of homoclinic orbits associated with the PtoP connection between $\Gamma_{*}$ and $\Gamma^{+}_1$, specifically examples of solution profiles and the associated bifurcation diagram. The $u_1$-traces of the two PtoP connections from Fig.~\ref{fig:PtoPphase} are shown in Figure~\ref{fig:HomGamma*}(a), which illustrates further how they connect $\Gamma_{*}$ and $\Gamma^{+}_1$. Notice the intermediate minimum in $u_1$ in panel~(a2), which corresponds to the loop around the additional orbit $^{A}\Gamma^{+}_1$ in Fig.~\ref{fig:PtoPphase}(b). Since both $\Gamma_*$ and $\Gamma^+_1$ are $R_1$-symmetric, the $R_1$-counterparts of the PtoP connections in Figure~\ref{fig:HomGamma*}(a), obtained by reflection in the $u_1$-axis, also exist: they correspond to return connections from $\Gamma^{+}_1$ to $\Gamma_{*}$. Hence, the two connections shown in panels (a) and their corresponding $R_1$-counterparts create different heteroclinic cycles between $\Gamma_{*}$ and $\Gamma^{+}_1$, and these organize different types of homoclinic solutions to $\Gamma_{*}$ and to $\Gamma^{+}_1$, respectively. 

Panels~(b) and~(c) of Fig.~\ref{fig:HomGamma*} show examples of homoclinic orbits to $\Gamma_{*}$ that are generated by PtoP cycles consisting of the two PtoP connections in Fig.~\ref{fig:HomGamma*}(a) and their $R_1$-counterparts. More specifically, the homoclinic solutions in panels~(b1) and~(c1) follow the PtoP connection in panel~(a1) from $\Gamma_{*}$ to near $\Gamma^{+}_1$ and then follow its $R_1$-counterpart back to $\Gamma_{*}$. Those in panels~(b2) and~(c2) follow, similarly, the PtoP connection in panel~(a2) and its $R_1$-counterpart. Finally, panels~(b3) and~(c3) show non-symmetric homoclinic solutions to $\Gamma_{*}$, which follow the PtoP connection in panel~(a1) and the $R_1$-counterpart of that in panel~(a2). The difference between panels~(b) and~(c) is that these homoclinic solutions make one and two loops around $\Gamma^{+}_1$, respectively. Indeed, by following the respective PtoP connections longer, homoclinic solutions to $\Gamma_{*}$ with any number of loops around $\Gamma^{+}_1$ can be found. 

Figure~\ref{fig:HomGamma*}(d) shows the one-parameter bifurcation diagram in $\beta_2$ of the PtoP connections, as well as the $R_1$-symmetric and non-symmetric homoclinic solutions to $\Gamma_{*}$, represented by the $L_2$-norm $||u_{1}||^2$ of a suitably truncated portion of the solutions. As in Fig.~\ref{fig:bifEtoP}, dashed vertical lines indicate the bifurcations  \textbf{BD} and \textbf{HH} that bound the $\beta_2$-range in region II where the equilibrium $\mathbf{0}$ is a saddle-focus. As is the case for the EtoP connections, the two PtoP connections in panels~(a1) and~(a2) lie on branches that emerge from \textbf{BD} and meet at a fold point to form a single curve. This fold, which we refer to as \textbf{PtoP}, occurs at $\beta_{2} \approx 0.5511$ and this is also indicated by a vertical dashed line in Fig.~\ref{fig:HomGamma*}(d); the PtoP connections in panels~(a1) and~(a2) are from the upper and the lower branch of this curve, respectively, and they no longer exist to the right of the line \textbf{PtoP}. We remark that the PtoP connections, which actually have an infinite $L_2$-norm, are represented in panel~(d) by the $L_2$-norm of their truncation with ten loops near each periodic solution $\Gamma_{*}$ and $\Gamma^{+}_1$. 

Also shown in Fig.~\ref{fig:HomGamma*}(d) are the four branches of $R_1$-symmetric and non-symmetric homoclinic solutions to $\Gamma_{*}$ with up to four oscillations near $\Gamma^{+}_1$. Since homoclinic connections to periodic orbits also have infinite $L_2$-norm, they are also represented here by the $L_2$-norm of a truncation, with four oscillations near $\Gamma_{*}$ in this case. As was the case for branches of homoclinic orbits to $\mathbf{0}$ in Sec.~\ref{sec:EtoP}, the different branches of increasing norm connect in pairs at fold bifurcation. More specifically, the $R_1$-symmetric homoclinic solutions to $\Gamma_{*}$ lie on branches that meet at fold points to form smooth curves. Here, the upper and lower branches represent homoclinic solutions generated by the PtoP connection in panel (a1) and its $R_1$-counterpart, and that in panel (a2) and its $R_1$-counterpart, respectively. The example solutions in panels (b1),(c1) and (b2),(c2) are indicated in Fig.~\ref{fig:HomGamma*}(d) by dots and crosses on the correspondingly colored branches. Similarly, the non-symmetric homoclinic solutions to $\Gamma_{*}$ lie on branches that meet at fold points, which are also the fold points of the corresponding $R_1$-symmetric solutions with the same number of loops around $\Gamma^{+}_1$. Hence, these folds are symmetry-breaking bifurcations of homoclinic orbits to $\Gamma_{*}$, which is why we refer to this family as $R_1$-symmetry-broken. Note that the pairs of branches of symmetry-broken solutions are not distinguished by the norm and appear to be on top of each other in Fig.~\ref{fig:HomGamma*}(d); the example solutions in panels (b3),(c3) are indicated in by diamonds on the correspondingly colored branches. As the number of loops around $\Gamma^{+}_1$ increases, the folds of the respective curves accumulate on the fold of PtoP connections from the left; notice that all these fold points have $\beta_2$-values that are very close to one another: in fact, they agree up to four decimal places with the value $\beta_{2}\approx0.5511$ of \textbf{PtoP}.

\subsection{Homoclinic orbits to $\Gamma_{1}^{+}$ }
\label{sec:HomGamma1+}

\begin{figure*}[t!]
   \centering
   \includegraphics{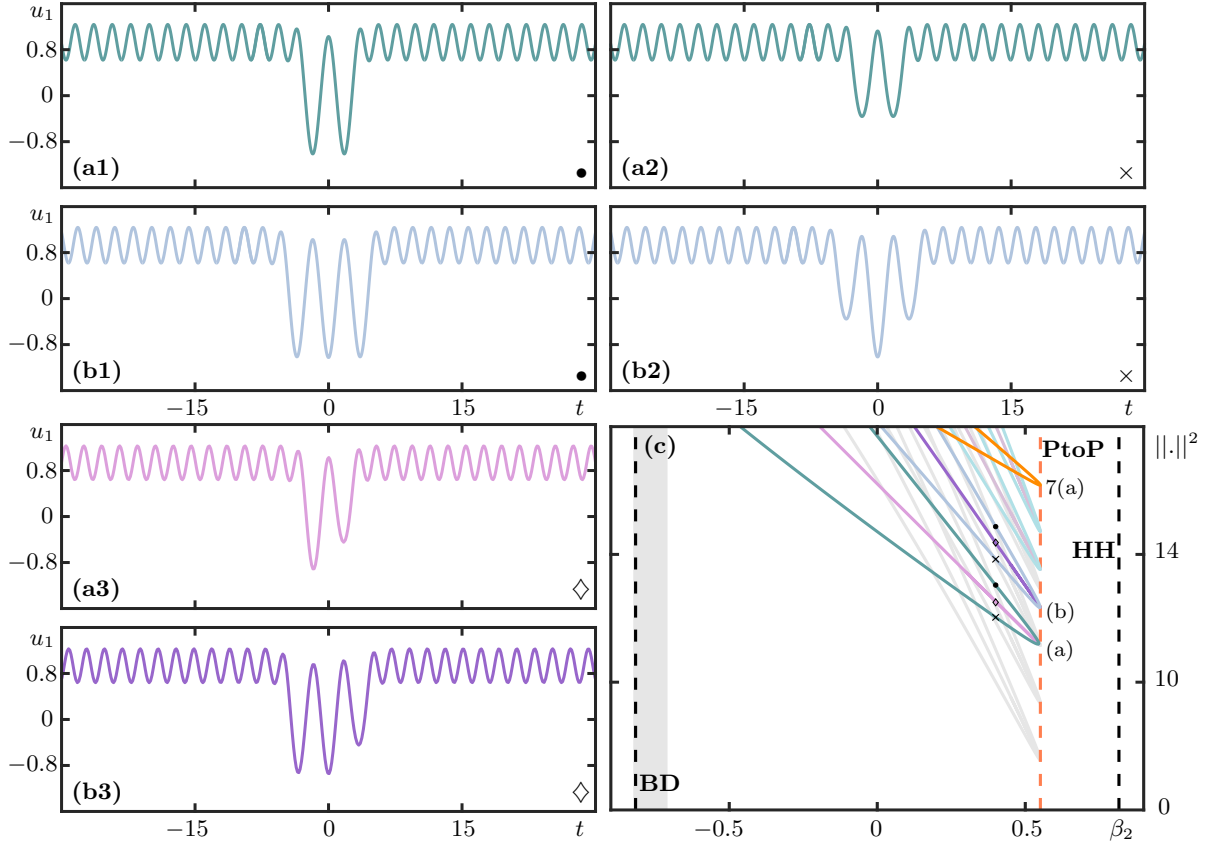} 
   \caption{Families of homoclinic solutions to $\Gamma^{+}_1$ associated with the PtoP connections from Fig.~\ref{fig:HomGamma*}(a). Panels (a1),(b1) and (a2),(b2) show $R_1$-symmetric, and panels (a3) and (b3) $R_1$-symmetry-broken homoclinic solutions to $\Gamma^{+}_1$. Panel (c) is the corresponding one-parameter bifurcation diagram in $\beta_2$ with the curve of PtoP connections (orange curve) and curves of homoclinic connections to $\Gamma^{+}_1$. The vertical dashed lines indicate \textbf{BD}, \textbf{HH} and the fold point \textbf{PtoP}; the color of the bifurcation curves and the dots, crosses and diamonds indicate the locations of the solutions in panels (a1) and (b1), (a2) and (b2), and (a3) and (b3), respectively.}  
      \label{fig:HomGamma1+}
\end{figure*} 

PtoP cycles organize homoclinic solutions to any periodic solutions involved in the cycle. In particular, and as Fig.~\ref{fig:HomGamma1+} shows, the pair of PtoP connections from Figs.~\ref{fig:PtoPphase} and~\ref{fig:HomGamma*}(a) also generates families of homoclinic solutions to the periodic solution $\Gamma^{+}_1$. Example solutions with one and two oscillations near $\Gamma_{*}$ are shown in Fig.~\ref{fig:HomGamma1+}(a) and (b), and panel~(c) is the associated bifurcation diagram that also includes the curve of PtoP connections. Comparison with Fig.~\ref{fig:HomGamma*}(b)--(d) shows that the different families of $R_1$-symmetric and $R_1$-symmetry-broken homoclinic solutions are organized in the same way. Specifically, Fig.~\ref{fig:HomGamma1+}(a1) and~(b1) show $R_1$-symmetric homoclinic solutions that follow the $R_1$-counterpart of the PtoP connection in Fig.~\ref{fig:HomGamma*}(a1) from $\Gamma^{+}_1$ to near $\Gamma_{*}$ and then follow this PtoP connection back to $\Gamma_{*}$. Figure~\ref{fig:HomGamma1+}(a2) and~(b2) show the same type of homoclinic orbit, but now for the other PtoP connection, and panels~(a3) and~(b3) show the corresponding $R_1$-symmetry-broken homoclinic solutions obtained from combining the two PtoP connections. The respective pairs of branches of homoclinic orbits with any number of loops around $\Gamma_{*}$, of which the first four are shown in Fig.~\ref{fig:HomGamma1+}(c), meet at fold points to form smooth curves; the locations of the different example solutions are again marked. The folds of the respective pairs coincide and are symmetry-breaking bifurcations of the homoclinic solutions to $\Gamma^{+}_1$. As the number of loops around $\Gamma_{*}$ increases, these fold points converge to the fold of the curve of PtoP connections; their $\beta_2$-values are indistinguishable (up to the accuracy of the continuation) from the value $\beta_{2}\approx0.5511$ of the point \textbf{PtoP}.

\subsection{\textbf{BD}-truncated snaking of generalized solitons}\label{sec:BDgeneral}

Figures~\ref{fig:HomGamma*}(d) and~\ref{fig:HomGamma1+}(c) show how the PtoP connections from Figures~\ref{fig:PtoPphase} organize families of $R_1$-symmetric and $R_1$-symmetry-broken homoclinic solutions to either $\Gamma^{*}$ and $\Gamma^{+}_1$. For increasing $\beta_2$, the curves of all these homoclinic solutions have folds, effectively all at the point \textbf{PtoP}, and disappear before reaching the bifurcation \textbf{HH}. Continuing these branches of homoclinic solutions to periodic orbits for decreasing $\beta_2$ turns out to be numerically more challenging. They have been computed to considerably lower values than shown in Figs.~\ref{fig:HomGamma*}(d) and~\ref{fig:HomGamma1+}(c), where their $L_2$-norm increases significantly for decreasing $\beta_2$. Nevertheless, our numerical computations strongly suggest that the homoclinic solutions to $\Gamma_{*}$ and $\Gamma^{+}_1$ are all also created at the Belyakov-Devaney bifurcation \textbf{BD}; however, this limit is not quite reached in the computation of the branches due to increased sensitivity of the continuations, which is arguably due to the oscillating nature of their tails.

\begin{figure*}[t!]
   \centering
   \includegraphics{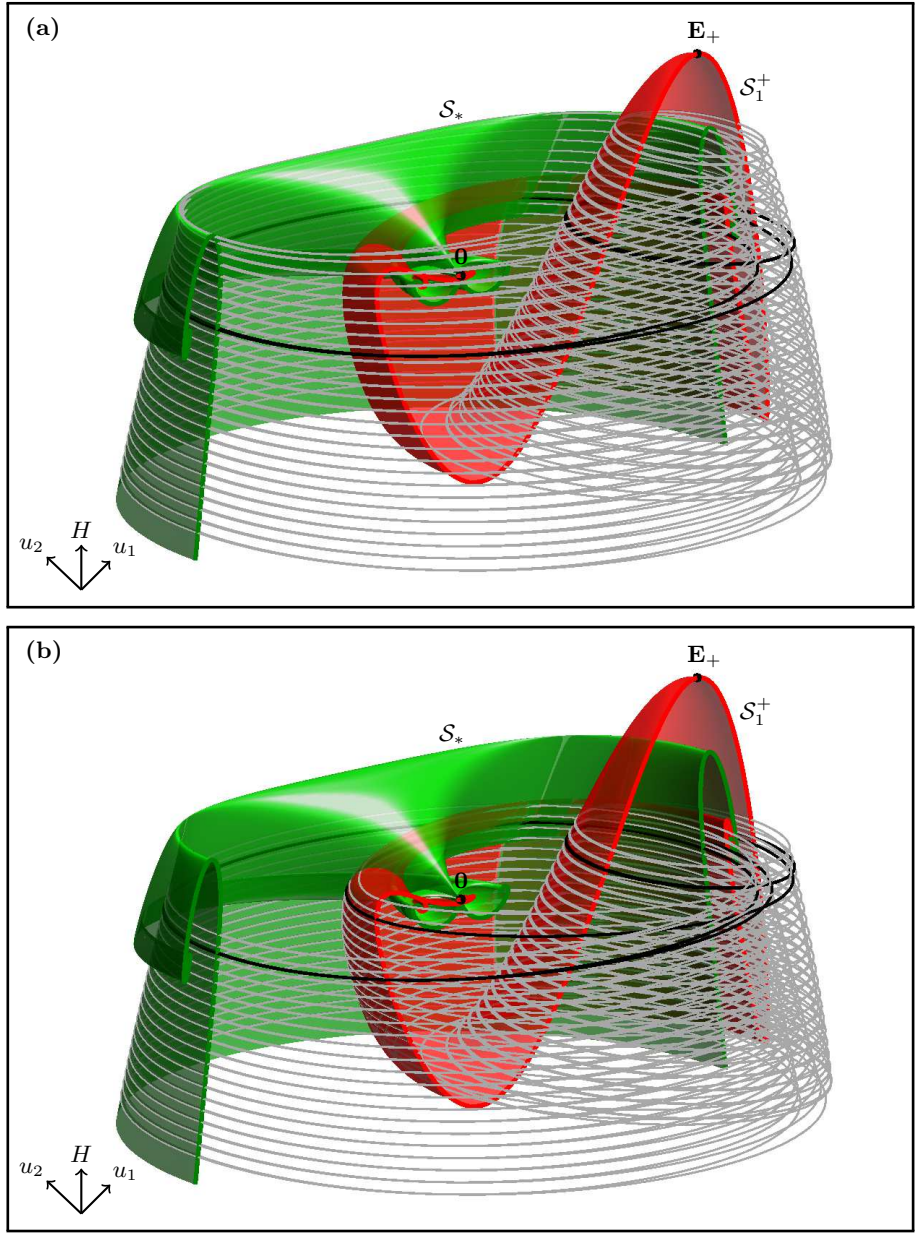} 
   \caption{Pair of surfaces in $(u_1,u_2,H)$-space of PtoP connections of system \eqref{eq:ODEsystem} for $\beta_2= 0.4$, shown with cutaway views of the surfaces $\mathcal{S}_*$ (green) and $\mathcal{S}_1^{+}$ (red) of periodic orbits. Specifically, panels~(a) and~(b) show a representative number of PtoP connections (gray curves), equally distributed in $H$, of the respective one-parameter family obtained by continuation of the PtoP connections in the zero-energy surface (black curves) from Fig.~\ref{fig:PtoPphase}(a) and~(b).} 
      \label{fig:PtoPsurfaces}
\end{figure*} 

Note that the $R_2$-symmetric counterparts of the homoclinic solutions in Fig.~\ref{fig:HomGamma*}(b1)--(c3) are also homoclinic solutions to $\Gamma_{*}$; however, they now make multiple loops around $\Gamma^{-}_1$. Since they have the same $L_2$-norm, their curves in the one-parameter bifurcation diagram are identical to those shown in panel (d). The $R_2$-symmetric counterparts of the homoclinic solutions to $\Gamma^{+}_1$ in Fig.~\ref{fig:HomGamma1+}(a1)--(b3) are simply the corresponding homoclinic solutions to $\Gamma^{-}_1$. Again, these have the same $L_2$-norm and so occur along the curves already shown in the bifurcation diagram in panel (c); hence, each curve represents homoclinic solutions both to $\Gamma^{ +}_1$ and to $\Gamma^{-}_1$.   

All these families of homoclinic orbits to periodic orbits of the ODE system~(\ref{eq:ODEsystem}) are generalized solitons (with non-decaying tails) of the GNLSE~\eqref{eq:gnlse}. In other words, as for the regular solitons with decaying tails \cite{PhysRevA.103.063514}, we find that the generalized solitons are also organized by the phenomenon of BD-truncated homoclinic snaking, where the homoclinic orbits are now to periodic orbits rather than to the equilibrium $\mathbf{0}$; compare Figs.~\ref{fig:HomGamma*}(d) and~\ref{fig:HomGamma1+}(c) with Fig.~\ref{fig:bifEtoP}. An important difference is that BD-truncated homoclinic snaking to periodic orbits exists not only in the zero-energy level, as we will show next. 

\subsection{Surfaces of PtoP connections}
\label{sec:PtoPsurfaces}

Since $\Gamma_{*}$ and $\Gamma^{+}_1$ lie in the zero-energy level, so do the PtoP connections in Fig.~\ref{fig:PtoPphase}. However, the periodic orbits can be continued to obtain the surfaces $\mathcal{S}_*$ and $\mathcal{S}_1^{+}$ from Fig.~\ref{fig:surfaces}(c) and (d). As Fig.~\ref{fig:PtoPsurfaces} shows, each of the two PtoP connections can similarly be continued in the energy $H$ to PtoP connections of corresponding periodic orbits in the respective energy level of $\mathcal{S}_*$ and $\mathcal{S}_1^{+}$. Here, panels~(a) and~(b) show the surfaces $\mathcal{S}_*$ and $\mathcal{S}_1^{+}$ in $(u_1, u_2, H)$-space, together with a number of representative PtoP connections of the corresponding one-parameter family; these are not rendered as a surface for clarity of presentation, and the PtoP connections from Fig.~\ref{fig:PtoPphase} are highlighted. Note that one could also continue each homoclinic solution in Fig.~\ref{fig:HomGamma*} and Fig.~\ref{fig:HomGamma1+} to find their respective continuations in $H$. We rather continue here the two PtoP connections themselves because they organize the many homoclinic solutions to the periodic orbits in the surfaces $\mathcal{S}_*$ and $\mathcal{S}_1^{+}$. In this way, we present a global picture that allows one to make connections to theoretical results on the existence of surfaces of homoclinic solutions to periodic solutions in reversible systems \cite{homburg_lamb_2006}. 

Figure~\ref{fig:PtoPsurfaces} clearly shows that the PtoP cycles and associated homoclinic orbits can be found over a range of the energy $H$; specifically, over the range of the joint existence of the continuations of the periodic orbits $\Gamma_{*}$ and $\Gamma^{+}_1$. As we continue the PtoP connections in the zero-energy level (black curves) towards positive and negative values of $H$, they reach a maximum and a minimum energy level, respectively, which correspond to folds with respect to $H$ of periodic orbits on the surfaces $\mathcal{S}_*$ and $\mathcal{S}_1^{+}$. We remark that these continuations are quite delicate near such local extrema of the energy, and it is difficult to continue a branch past such a fold. Nevertheless, we are able to identify the respective PtoP connection past the fold and resume the continuation in this way. More specifically, as we continue the PtoP connection in Fig.~\ref{fig:PtoPphase}(a) towards the positive values of $H$, it reaches the energy level with $H\approx3.3$, which is the global maximum of $\mathcal{S}_*$; see Fig.~\ref{fig:PtoPsurfaces}(a). When we continue it towards negative values of $H$, it reaches the energy level with $H\approx-6.2$, which is the global minimum of $\mathcal{S}_1^{+}$. Likewise, when the connection in Fig.~\ref{fig:PtoPphase}(b) is continued towards negative values of $H$, it also reaches the energy level with $H\approx-6.2$; see Fig.~\ref{fig:PtoPsurfaces}(b). However, as we continue it towards positive values of $H$, it only reaches the energy level with $H\approx0.29$. This is because this family of PtoP connections has an extra loop near the continuation of the $R_1$-symmetric periodic orbit $^{A}\Gamma^{+}_1$ on $\mathcal{S}_1^{+}$, which only exists up to the local maximum of $\mathcal{S}_1^{+}$ at energy level $H\approx0.29$. The close passage of the PtoP connection in Fig.~\ref{fig:PtoPphase}(b) of $^{A}\Gamma^{+}_1$ already demonstrates that it is possible to find PtoP connections between $\Gamma_{*}$ and $^{A}\Gamma^{+}_1$ and between $\Gamma^{+}_1$ and $^{A}\Gamma^{+}_1$. Overall, past any fold on the surface $\mathcal{S}_*$ or $\mathcal{S}_1^{+}$ the respective PtoP connection connects different periodic orbits, and they all exist over a certain range of $H$.

Any PtoP connections in a given energy level can be combined to create heteroclinic PtoP cycles between different periodic orbits and associated homoclinic orbits, that is, additional families of generalized solitons that exist over certain ranges of $H$ near $H=0$. In particular, there are infinitely many families of homoclinic orbits to the periodic orbits $\Gamma_{*}$ and $\Gamma^{\pm}_1$ in the zero-energy level. This geometric picture allows us to conclude the following. As $\Gamma_{*}$ and/or $\Gamma^{\pm}_1$ approach the primary homoclinic orbits in the zero-energy level, the families of the associated homoclinic orbits to the periodic orbits $\Gamma_{*}$ and/or  $\Gamma^{\pm}_1$ approach connecting orbits from and back to the primary homoclinic orbits to $\mathbf{0}$. Taking the limit may, hence, be a mechanism for creating homoclinic orbits to homoclinic orbits, which are also referred to as super-homoclinic orbits \cite{bakrani2021invariant}. These objects are limits of generalized solitons in the GNLSE, and it is an interesting question beyond the scope of this paper whether and how they they act as organizing centers for the different types of solitons.

\section{P{\scriptsize to}P connections and multi-oscillation solitons}
\label{sec:PtoPmulti}

In the zero-energy level one finds the PtoP connections between $\Gamma_{*}$ and $\Gamma^{+}_1$ from Fig.~\ref{fig:PtoPphase}, as well as their $R_1$- and $R_2$-counterparts that involve $\Gamma^{-}_1$. Together they are involved in creating more complicated heteroclinic PtoP cycles between the three periodic orbits $\Gamma_{*}$, $\Gamma^{+}_1$ and $\Gamma^{-}_1$. As we show now, these PtoP cycles can be combined with the EtoP connections from Sec.~\ref{sec:EtoP} effectively in any combination to create more complicated heteroclinic cycles from and back to $\mathbf{0}$. In turn, these organize a plethora of homoclinic orbits to $\mathbf{0}$, which all correspond to solitons of the GNLSE with decaying tails that feature episodes of oscillations near any of these periodic orbits. For this reason, we refer to them as multi-oscillation solitons.

\begin{figure}[t!]
   \centering
   \includegraphics[width=7.2cm]{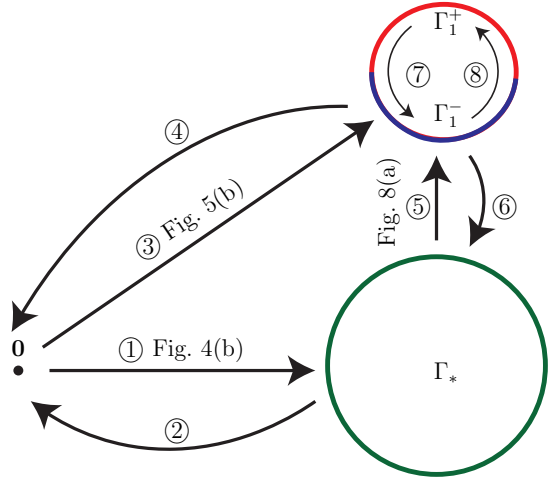} 
   \caption{Schematic digram showcasing all the possible connections between $\mathbf{0}$ (black dot), $\Gamma_{*}$ (green circle), and $\Gamma^{\pm}_1$ (red/blue circle). Arrows indicate different EtoP and PtoP connections with reference to previous figures, and they are labeled {\textcircled{\raisebox{-0.9pt}{1}}} to {\textcircled{\raisebox{-0.9pt}{8}}}.}  
      \label{fig:multisketch}
\end{figure} 

This whole mechanism is summarized in Fig.~\ref{fig:multisketch}, which shows a schematic diagram in the form of a graph of possible connections between $\mathbf{0}$, $\Gamma_{*}$, and $\Gamma^{\pm}$ at $\beta_2$-values where all of them exist, such as for $\beta =0.4$ as we used throughout. As we discussed in Sec~\ref{sec:EtoP} and Sec.~\ref{sec:PtoPgeneral}, there exist two distinct EtoP and PtoP connections for fixed parameter values, and they are represented in Fig.~\ref{fig:multisketch} by single directed edges that are numbered {\textcircled{\raisebox{-0.9pt}{1}}} to {\textcircled{\raisebox{-0.9pt}{8}}}; here the arrows indicate the direction of time and reference to previous figures indicates already the encountered connections. The edges labeled {\textcircled{\raisebox{-0.9pt}{7}}} and {\textcircled{\raisebox{-0.9pt}{8}}} represent pairs of PtoP connections between the $R_2$-counterparts $\Gamma^{+}_1$ and $\Gamma^{-}_1$, which are jointly represented by a single bi-colored circle.  Furthermore, in this labeling, the even connections \textcircled{\raisebox{-0.9pt}{2}}, \textcircled{\raisebox{-0.9pt}{4}}, \textcircled{\raisebox{-0.9pt}{6}} and \textcircled{\raisebox{-0.9pt}{8}} correspond to the $R_1$- or $R_2$-counterparts of the odd connections \textcircled{\raisebox{-0.9pt}{1}}, \textcircled{\raisebox{-0.9pt}{3}}, \textcircled{\raisebox{-0.9pt}{5}} and \textcircled{\raisebox{-0.9pt}{7}}, respectively. 

Any path in the directed graph in Fig.~\ref{fig:multisketch} that starts and ends at $\mathbf{0}$ represents certain heteroclinic cycles with associated infinite families of homoclinic orbits to $\mathbf{0}$ of the different symmetry types, which are all solitons of the GNLSE. Notice that the families of homoclinic solutions that were previously found \cite{PhysRevA.103.063514} and are summarized in Sec.~\ref{sec:EtoP} are organized by the cycles that are formed by only the connections \textcircled{\raisebox{-0.9pt}{1}} and \textcircled{\raisebox{-0.9pt}{2}}, and by \textcircled{\raisebox{-0.9pt}{3}} and \textcircled{\raisebox{-0.9pt}{4}}. That is, they reach $\Gamma_{*}$ or $\Gamma^{+}_1$ along the EtoP connection  \textcircled{\raisebox{-0.9pt}{1}} or \textcircled{\raisebox{-0.9pt}{3}}, loop multiple type times close to the respective periodic orbit, and then converge back to $\mathbf{0}$ along the EtoP connection \textcircled{\raisebox{-0.9pt}{2}} or \textcircled{\raisebox{-0.9pt}{4}}. Importantly, the existence of the PtoP connections \textcircled{\raisebox{-0.9pt}{5}} to \textcircled{\raisebox{-0.9pt}{8}} that connect the periodic orbits $\Gamma_{*}$ and $\Gamma^{\pm}_1$ allows for more complicated homoclinic solutions to $\mathbf{0}$ to be constructed. Each arrow represents two distinct connections from one saddle object to another, which makes this mechanism more complex than Fig~\ref{fig:multisketch} suggests. 

Our numerical evidence supports the observation/conjecture that for any path from $\mathbf{0}$ back to itself in Fig~\ref{fig:multisketch}, the corresponding EtoP cycles with associated families of homoclinic solutions to $\mathbf{0}$ indeed exist in system~(\ref{eq:ODEsystem}). To illustrate this statement, we consider two specific paths and compute and show the corresponding families of homoclinic solutions. We first consider the heteroclinic cycles given by the edge sequence
\begin{equation*}
\mathbf{0} \xrightarrow{\textrm{\textcircled{\raisebox{-0.5pt}{1}}}}~\Gamma^{*}~\xrightarrow{\textrm{\textcircled{\raisebox{-0.5pt}{5}}}}~{\Gamma^{+}_1}~\xrightarrow{\textrm{\textcircled{\raisebox{-0.5pt}{6}}}}~\Gamma^{*}~\xrightarrow{\textrm{\textcircled{\raisebox{-0.5pt}{2}}}}~\mathbf{0},
\end{equation*}
which we say is of type (\textcircled{\raisebox{-0.9pt}{1}}, \textcircled{\raisebox{-0.9pt}{5}}, \textcircled{\raisebox{-0.9pt}{6}}, \textcircled{\raisebox{-0.9pt}{2}}). Since the periodic orbit $\Gamma^{+}_1$ in the center is only $R_1$-symmetric, this cycle only organizes families of $R_1$-symmetric and $R_1$-symmetry-broken homoclinic solutions. 

Secondly, we consider the heteroclinic cycles given by the edge sequence
\begin{equation*}
\mathbf{0} \xrightarrow{\textrm{\textcircled{\raisebox{-0.5pt}{3}}}}~\Gamma^{+}_1~\xrightarrow{\textrm{\textcircled{\raisebox{-0.5pt}{6}}}}~{\Gamma_{*}}~\xrightarrow{\textrm{\textcircled{\raisebox{-0.5pt}{5}}}}~\Gamma^{+}_1/\Gamma^{-}_1~\xrightarrow{\textrm{\textcircled{\raisebox{-0.5pt}{4}}}}~\mathbf{0},
\end{equation*}
which are of type (\textcircled{\raisebox{-0.9pt}{3}}, \textcircled{\raisebox{-0.9pt}{6}}, \textcircled{\raisebox{-0.9pt}{5}}, \textcircled{\raisebox{-0.9pt}{4}}). Since now the periodic orbit $\Gamma_{*}$ at the center is $R_*$-symmetric, this cycle organizes families of both $R_1$- and $R_2$-symmetric homoclinic solutions and the corresponding symmetry-broken homoclinic solutions.  

For each of these two types of more complex homoclinic solutions to $\mathbf{0}$, we present again representative examples as well as the bifurcation diagram in $\beta_2$ with the corresponding branches. 

\subsection{Solitons of type (\textcircled{\raisebox{-0.9pt}{1}}, \textcircled{\raisebox{-0.9pt}{5}}, \textcircled{\raisebox{-0.9pt}{6}}, \textcircled{\raisebox{-0.9pt}{2}})}
\label{sec:multiFirsttype}

\begin{figure*}[t!]
   \centering
   \includegraphics{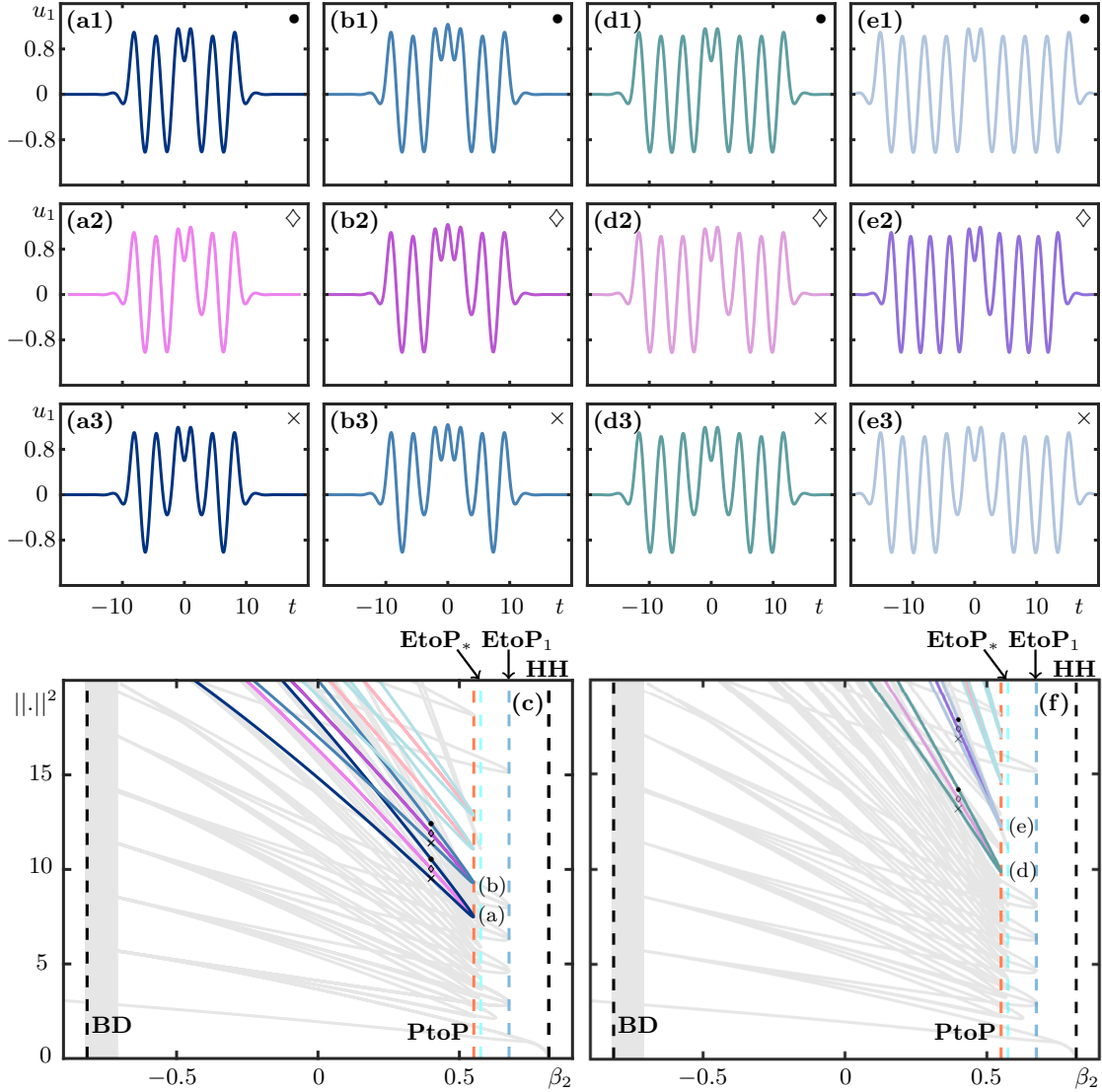} 
   \caption{Families of homoclinic solutions to $\mathbf{0}$ of system \eqref{eq:ODEsystem} generated by the EtoP and PtoP connections of the edge sequence ({\textcircled{\raisebox{-0.9pt}{1}}, \textcircled{\raisebox{-0.9pt}{5}}, \textcircled{\raisebox{-0.9pt}{6}}, \textcircled{\raisebox{-0.9pt}{2}}}). Panels (a1),(b1) and (d1),(e1) and (a3),(b3) and (d3),(e3) show $R_1$-symmetric, and panels (a2),(b2) and (d2),(e2) $R_1$-symmetry-broken homoclinic solutions, all for $\beta_2=0.4$. Panels (c) and (f) are the bifurcation diagrams in $\beta_2$ of the homoclinic solutions in panels (a) and (b), and of those in panels (d) and (e), respectively. The grey curves are the branches from Figure~\ref{fig:bifEtoP}. The vertical dashed lines indicate \textbf{BD}, \textbf{HH} and the fold points \textbf{EtoP$_*$}, \textbf{EtoP$_1$} and \textbf{PtoP}; the color of the bifurcation curves and the dots, crosses and diamonds indicate the locations of the solutions in panels (a),(b) and (d),(f), respectively.}  
      \label{fig:multiFirstseq}
\end{figure*} 

Figure~\ref{fig:multiFirstseq} shows families of homoclinic solutions to $\mathbf{0}$ for $\beta_2 = 0.4$ that are organized by the heteroclinic cycles with the edge sequence (\textcircled{\raisebox{-0.9pt}{1}}, \textcircled{\raisebox{-0.9pt}{5}}, \textcircled{\raisebox{-0.9pt}{6}}, \textcircled{\raisebox{-0.9pt}{2}}). This type is determined by the fact that the connections \textcircled{\raisebox{-0.9pt}{6}} and \textcircled{\raisebox{-0.9pt}{2}} are the $R_1$-counterparts of the connections \textcircled{\raisebox{-0.9pt}{5}} and \textcircled{\raisebox{-0.9pt}{1}}, respectively. As panels (a),(b) and (d),(f) of Fig.~\ref{fig:multiFirstseq} show, multi-oscillation solitons of this type correspond to homoclinic orbits that leave $\mathbf{0}$ to oscillate around $\Gamma_{*}$, transition to oscillate around $\Gamma^{+}_1$, transition back and oscillate around $\Gamma_{*}$ again, and finally return to $\mathbf{0}$. The connections can be made by either of the pair of EtoP from Fig.~\ref{fig:solG*}(a) and PtoP connections from Fig.~\ref{fig:HomGamma*}(a), and the number of oscillations of each of these episodes differs from branch to branch, as is illustrated in the bifurcation diagram in Fig.~\ref{fig:multiFirstseq}(c) and (f).

More specifically, Fig.~\ref{fig:multiFirstseq}(a1) shows an $R_1$-symmetric homoclinic solutions of this type: from $\mathbf{0}$ it follow the EtoP connection in Fig.~\ref{fig:solG1pm}(a2); makes two loops close to $\Gamma_{*}$; follows the PtoP connection in Fig.~\ref{fig:HomGamma*}(a1); makes one loop close to $\Gamma^{+}_1$; follows the $R_1$-counterpart of the PtoP connection in Fig.~\ref{fig:HomGamma*}(a1); makes two loops close to $\Gamma_{*}$; and finally follows the $R_1$-counterpart of the EtoP connection in Fig.~\ref{fig:solG1pm}(a2) back to $\mathbf{0}$. The homoclinic solution in panel (a3) of Fig.~\ref{fig:multiFirstseq} differs only in that it follows the PtoP connection in Fig.~\ref{fig:HomGamma*}(a2) and its $R_1$-counterpart. The homoclinic solution in panel (a2), on the other hand, is $R_1$-symmetry-broken, because it first follows the EtoP and PtoP connection in Fig.~\ref{fig:solG1pm}(a2) and Fig.~\ref{fig:HomGamma*}(a1), respectively, and then converges back to $\mathbf{0}$ by following the $R_1$-counterparts of the PtoP and EtoP connections in Fig.~\ref{fig:HomGamma*}(a2) and Fig.~\ref{fig:solG1pm}(a2), respectively. Fixing the number of loops near $\Gamma_{*}$ and increasing the number of loops near $\Gamma^{+}_1$ generates families of $R_1$-symmetric and $R_1$-symmetry-broken homoclinic solutions, and panels (b1), (b2) and (b3) show the ones with one additional loop.

Figure~\ref{fig:multiFirstseq}(c) shows the bifurcation diagram in $\beta_2$ of these families, where solutions are again represented by their $L_2$-norm $||.||^2$. Here, we show the curves of the homoclinic solutions with two to five loops near $\Gamma^{+}_1$; the bifurcations \textbf{BD} and \textbf{HH} that bound region II are shown by dashed vertical lines, as are the fold points \textbf{EtoP$_*$}, \textbf{EtoP$_1$} and \textbf{PtoP}. For reference, all previously found branches of homoclinic solutions\cite{PhysRevA.103.063514}  from Sec.~\ref{sec:EtoP} are shown in light gray. The two $R_1$-symmetric homoclinic solutions in panels (a1) and (a3) of Fig.~\ref{fig:multiFirstseq}, with the same number of loops near $\Gamma_{*}$ and $\Gamma^{+}_1$, respectively, lie on branches that connect at a fold point to form a smooth curve. As is the case for the simpler homoclinic orbits, this fold is also a point of symmetry breaking from which bifurcates a pair of branches of $R_1$-symmetry-broken homoclinic solutions; the one in panel (a2) lies on one of these branches (which coincide because they have the same $L_2$-norm). This picture of branches connecting as pairs at joint fold points is repeated in panel~(c) when the number of loops near $\Gamma^{+}_1$ is increased; the positions of the homoclinic solutions in panels (b) are indicated on the respective branches.

\begin{figure*}[t!]
   \centering
   \includegraphics{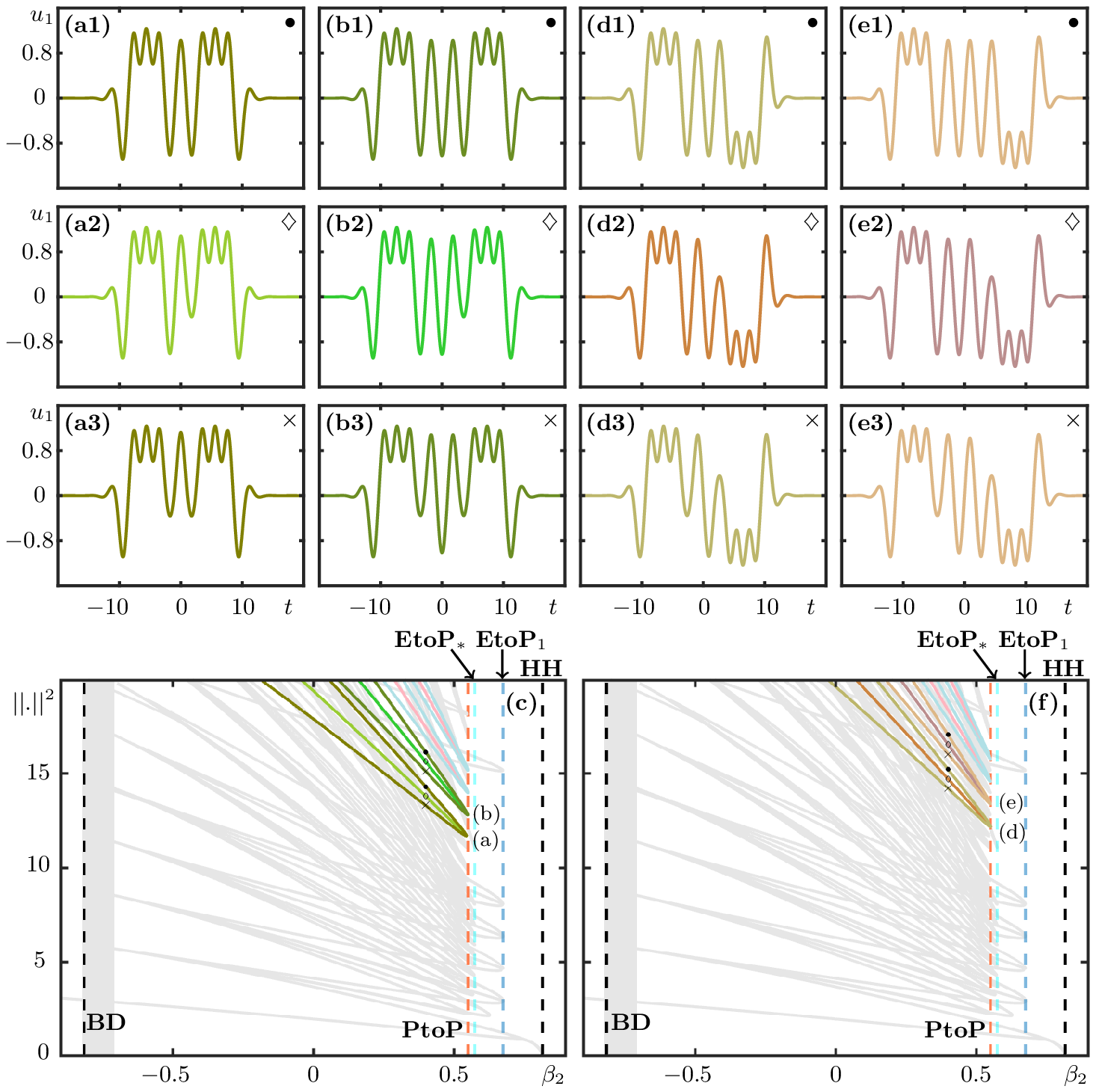} 
   \caption{Families of homoclinic solutions to $\mathbf{0}$ of system \eqref{eq:ODEsystem} generated by the EtoP and PtoP connections of the edge sequence ({\textcircled{\raisebox{-0.9pt}{3}}, \textcircled{\raisebox{-0.9pt}{6}}, \textcircled{\raisebox{-0.9pt}{5}}, \textcircled{\raisebox{-0.9pt}{4}}}). Panels (a1),(b1) and (a3),(b3) show $R_1$-symmetric, panels (a2),(b2) $R_1$-symmetry-broken, panels (d1),(e1) and (d3),(e3) show $R_2$-symmetric, and panels (d2),(e2) $R_2$-symmetry-broken homoclinic solutions, all for $\beta_2=0.4$. Panels (c) and (f) are the bifurcation diagrams in $\beta_2$ of the homoclinic solutions in panels (a) and (b), and of those in panels (d) and (e), respectively. The grey curves are the branches from Figure~\ref{fig:bifEtoP}. The vertical dashed lines indicate \textbf{BD}, \textbf{HH} and the fold points \textbf{EtoP$_*$}, \textbf{EtoP$_1$} and \textbf{PtoP}; the color of the bifurcation curves and the dots, crosses and diamonds indicate the locations of the solutions in panels (a),(b) and (d),(f).}  
      \label{fig:multiSecondseq}
\end{figure*} 

More families of homoclinic solutions associated with type  (\textcircled{\raisebox{-0.9pt}{1}}, \textcircled{\raisebox{-0.9pt}{5}}, \textcircled{\raisebox{-0.9pt}{6}}, \textcircled{\raisebox{-0.9pt}{2}}) can be generated by fixing the oscillations near $\Gamma^{+}_1$ and increasing the number of loops near $\Gamma_{*}$. As before, there exist $R_1$-symmetric and $R_1$-symmetry-broken families of homoclinic solutions, and examples of them are shown in panels (d) and (e) of Fig.~\ref{fig:multiFirstseq}. Notice that those in panels (d) are like the ones in panels (a), but with one further loop near $\Gamma_{*}$; similarly, the homoclinic solutions in panels (e) have two extra loops near $\Gamma_{*}$. As the bifurcation diagram in panel (f) shows, all these $R_1$-symmetric and $R_1$-symmetry-broken homoclinic solutions also lie on a single curve with two branches that meet at a fold point, from which branches of the $R_1$-symmetry-broken homoclinic solutions bifurcate. 

\subsection{Solitons of type (\textcircled{\raisebox{-0.9pt}{3}}, \textcircled{\raisebox{-0.9pt}{6}}, \textcircled{\raisebox{-0.9pt}{5}}, \textcircled{\raisebox{-0.9pt}{4}})}
\label{sec:multiSecondtype}

Figure~\ref{fig:multiSecondseq} shows families of homoclinic solutions to $\mathbf{0}$ that are organized by the heteroclinic cycles with the edge sequence (\textcircled{\raisebox{-0.9pt}{3}}, \textcircled{\raisebox{-0.9pt}{6}}, \textcircled{\raisebox{-0.9pt}{5}}, \textcircled{\raisebox{-0.9pt}{4}}). Shown are examples of $R_1$-symmetric and $R_1$-symmetry-broken homoclinic solutions in panels (a) and (b), and of $R_2$-symmetric and $R_2$-symmetry-broken homoclinic solutions in panels (a) and (b), respectively; the corresponding bifurcation diagrams in $\beta_2$ are shown in panels (c) and (f). For the $R_1$-symmetric and $R_1$-symmetry-broken homoclinic solutions, the connections \textcircled{\raisebox{-0.9pt}{5}} and \textcircled{\raisebox{-0.9pt}{4}} are the $R_1$-counterparts of the connections \textcircled{\raisebox{-0.9pt}{6}} and \textcircled{\raisebox{-0.9pt}{3}}, respectively. All these homoclinic solutions loop close to $\Gamma^{+}_1$ before converging back to $\mathbf{0}$. For the $R_2$-symmetric and $R_2$-symmetry-broken homoclinic solutions, the connections \textcircled{\raisebox{-0.9pt}{5}} and \textcircled{\raisebox{-0.9pt}{4}} are the $R_2$-counterparts of the connections \textcircled{\raisebox{-0.9pt}{6}} and \textcircled{\raisebox{-0.9pt}{3}}, respectively; so now all these homoclinic solutions loop close to $\Gamma^{-}_1$ before converging back to $\mathbf{0}$.  

To be more specific, the homoclinic solution in Fig.~\ref{fig:multiSecondseq}(a1) follows the EtoP connection in Fig.~\ref{fig:solG1pm}(a2) and makes two loops near $\Gamma^{+}_1$. After that, it follows the $R_1$-counterpart of the PtoP connection in Fig.~\ref{fig:HomGamma*}(a1) and makes one loop near $\Gamma_{*}$; subsequently, it follows the PtoP connection in Fig.~\ref{fig:HomGamma*}(a1) and the $R_1$-counterpart of the EtoP connection in Fig.~\ref{fig:solG1pm}(a1) to converge back to $\mathbf{0}$. Similarly, the homoclinic solution in panels (a3) also makes use of the EtoP connection in Fig.~\ref{fig:solG1pm}(a1); however, it is now associated with the PtoP connection in Fig.~\ref{fig:HomGamma*}(a2). The homoclinic solutions in panels (b1) and (b3) are similar to the ones in panels (a1) and (a3), but make one additional loop near $\Gamma_{*}$. The corresponding $R_1$-symmetry-broken homoclinic solutions are shown in panels (a2) and (b2), respectively. Similar to what we found in Fig.~\ref{fig:multiFirstseq}, one can also generate other families of homoclinic solutions by fixing the number of loops they make near $\Gamma_{*}$ and by increasing the number of loops near $\Gamma^{+}_1$. As the bifurcation diagram in Fig.~\ref{fig:multiSecondseq}(c) shows, all these  $R_1$-symmetric and $R_1$-symmetry-broken homoclinic solutions also lie on a single curve with two branches that meet at a fold point.

It is also possible to generate families of $R_2$-symmetric and $R_2$-symmetry-broken homoclinic solutions of type (\textcircled{\raisebox{-0.9pt}{3}}, \textcircled{\raisebox{-0.9pt}{6}}, \textcircled{\raisebox{-0.9pt}{5}}, \textcircled{\raisebox{-0.9pt}{4}}), and examples are shown in panels (d) and (e) of Fig.~\ref{fig:multiSecondseq}. As panels (d1),(e1) and (d3),(e3) show, such $R_2$-symmetric homoclinic solutions initially loop twice close to $\Gamma^{+}_1$, then follow the respective PtoP connections to loop close to $\Gamma_{*}$ and, finally, loop three times close to $\Gamma^{-}_1$, instead of $\Gamma^{+}_1$, before returning to $\mathbf{0}$. As before, the corresponding $R_2$-symmetry-broken homoclinic solutions can be constructed by combining the two different PtoP connections, and they are shown in panels (d2) and (e2). The difference between the homoclinic orbits in panels (d) and (e) is that they have one and two loops near $\Gamma_{*}$, respectively. Taking more loops near $\Gamma_{*}$ creates additional homoclinic orbits of the respective family, of which the first four are shown in the bifurcation diagram in Fig.~\ref{fig:multiSecondseq}(f). Also the $R_2$-symmetric homoclinic solutions lie in pairs on curves with two branches that meet at fold points, from which branches of the corresponding $R_2$-symmetry-broken homoclinic solutions emerge.

\subsection{\textbf{BD}-truncated snaking of multi-oscillation solitons}
\label{sec:BDmulti}

Figures~\ref{fig:multiFirstseq} and~\ref{fig:multiSecondseq} illustrate the existence of an entire menagerie of additional homoclinic orbits to $\mathbf{0}$ that visit any of the periodic orbits $\Gamma_{*}$ and $\Gamma_{1}^{\pm}$. These additional homoclinic solutions are indeed all multi-oscillation solitons of the GNLSE. More specifically, for any chosen edge sequence, one can find the associated families of symmetric and symmetry-broken homoclinic solutions; these include $R_2$-symmetric and $R_2$-symmetry-broken ones when the respective heteroclinic cycles is not $R_2$-symmetric itself. Figures~\ref{fig:multiFirstseq} and~\ref{fig:multiSecondseq} show certain families given by increasing only the number of loops near one of the constituent periodic orbits. However, we observe and conjecture that homoclinic solutions exist for any prescribed number of loops near any of the constituent periodic orbits of a given type. Note also that the $R_2$-counterparts of the shown homoclinic solutions also exist. We do not show them here because, on the level of the $u_1$-traces shown in these figures, they simply correspond to reflections of $u_1$ in the $t$-axis, so that maxima become minima and vice versa, and $\Gamma_{1}^{+}$ and $\Gamma_{1}^{-}$ are exchanged.

As the bifurcation diagrams in panels (c) and (f) of Figs.~\ref{fig:multiFirstseq} and~\ref{fig:multiSecondseq} illustrate,  all these multi-oscillation solitons lie on curves organized by BD-truncated homoclinic snaking. Indeed, in spite of increasing numerical sensitivity due to their more complicated nature, we are able to continue all branches of multi-oscillation solitons for decreasing $\beta_2$ to quite near the point \textbf{BD} to confirm this. All these solutions require the existence of the respective EtoP and PtoP connections. Since the fold \textbf{PtoP} of the latter occurs for smaller $\beta_2$ than the folds \textbf{EtoP$_*$}, \textbf{EtoP$_1$} of the different EtoP connections, all these multi-oscillation solitons involving the periodic orbits $\Gamma_{*}$ and $\Gamma_{1}^{\pm}$ lie on branches that have folds near \textbf{PtoP}. In fact, as was also the case for the folds of generalized solitons in Sec.~\ref{sec:PtoPgeneral}, the fold points of multi-oscillation solitons are indistinguishable in their $\beta_2$-value (as found by continuation) from that of the fold \textbf{PtoP}.

\section{Discussions and outlook}
\label{sec:conclusions}

We showed that there exist infinitely many infinite families of generalized and multi-oscillation solitons of the GNLSE, in addition to the families of `regular'  solitons that were found before \cite{PhysRevA.103.063514}. To achieve this, we used a dynamical system approach to translate solitons of the GNLSE into homoclinic orbits to the equilibrium $\mathbf{0}$ in the zero-energy level of the ODE system~\eqref{eq:ODEsystem}, which is Hamiltonian and features two reversible symmetries, $R_1$ and $R_2$. The new kinds of solitons crucially involve transitions between periodic solutions, which are given by PtoP connections between saddle-type periodic orbits of the ODE. For the specific example of the primary periodic orbits $\Gamma_{*}$ and $\Gamma_{1}^{\pm}$ in the zero-energy level, we showed how PtoP connections between them give rise to infinite families of homoclinic orbits to either of the constituent periodic orbits. In terms of the GNLSE, these solutions are generalized solitons with non-decaying oscillating tails; notably, these PtoP connections and associated generalized solitons are not confined to the zero-energy level. Furthermore, PtoP connections in the zero-energy level can be combined with EtoP connections from $\mathbf{0}$ to the respective periodic orbits into heteroclinic cycles. As we argued, these generate infinite families of homoclinic solutions to $\mathbf{0}$, which are multi-oscillation solitons of the GNLSE with different episodes of oscillations near any of the periodic orbits they visit. We stress that these solitons are not bound states or `molecules' of two or more interacting primary solitons. 

All these solitons can be continued in parameters. Specifically, we showed by changing the strength $\beta_2$ of the quadratic dispersion that all solitons are organized as BD-truncated homoclinic snaking, even the generalized solitons that correspond to homoclinic solutions to periodic orbits. This means that, for all families of solitons, pairs of branches start at the Belyakov-Daveney bifurcation \textbf{BD} and meet at fold bifurcations that are also symmetry-breaking bifurcation of the respective solutions. The folds of a given family accumulate on the folds of the respective EtoP or ProP connections that are organizing it. Note, in particular, that all these solutions of the GNLSE exist for $\beta_2 = 0$, that is, for the case of pure quartic dispersion.

For heteroclinic cycles involving only the equilibrium $\mathbf{0}$ and the primary periodic orbits $\Gamma_{*}$ and $\Gamma_{1}^{\pm}$, we presented a directed graph with numbered edges to encode the possibilities: any edge sequence in this graph from and back to $\mathbf{0}$ generates associated families of multi-oscillation solitons of the respective type. While we supported it only for certain families, it is a natural conjecture that this statement generally holds. What is more, the graph can also be used to construct more complicated generalized solitons by selecting edge sequences of this graph that avoid the equilibrium. However, the story is even more complicated: already the basic surfaces $\mathcal{S}_{*}$ and $\mathcal{S}_{1}^{\pm}$ intersect the zero-energy level infinitely many times as they converge onto the (union of the) primary homoclinic orbits. Hence, the graph we presented can be extended with infinitely more periodic orbits of $\mathcal{S}_{*}$ and $\mathcal{S}_{1}^{\pm}$ and associated edges that represent EtoP connections from $\mathbf{0}$ to them, as well as PtoP connections between all these additonal periodic orbits. We conjecture that any edge sequence in this much extended graph also generates families of multi-oscillation solitons of the corresponding type, as was shown for the cases presented here. But there is even more: any of the thus created solitons is itself the limit of a family of periodic orbits! We suspect that each of them, when continued in the energy, creates infinitely many periodic orbits in the zero-energy level that also feature EtoP and PtoP connections, and so on. The picture that emerges is truly a never-ending and `self-similar' plethora of additional solitons, which we conjecture are all created at the point \textbf{BD} as part of the overall phenomenon of BD-truncated homoclinic snaking of the ODE system~\eqref{eq:ODEsystem} and, hence the GNLSE~\eqref{eq:gnlse}. More generally, we conjecture that a Belyakov-Devaney bifurcation in any fourth-order Hamiltonian system with two reversible symmetries is an organizing center near which all of the solutions reported here can be found in the same way.

There  are certainly avenues for future research that arise from our study. First of all, we already alluded to the geometry of the basic surfaces $\mathcal{S}_{*}$ and $\mathcal{S}_{1}^{\pm}$, which we only considered here for the fixed parameter value $\beta_2 = 0.4$. As $\beta_2$ is increased, these surfaces interact and their different parts connect differently in a sequence of transitions through specific types of symmetry-breaking bifurcations, period-$k$ multiplying bifurcations, and saddle-node bifurcations; these results will be presented elsewhere \cite{bgbk_periodic}. Secondly, it remains a considerable challenge to formally prove the existence of the EtoP and PtoP connections and all the different families of homoclinic orbits they generate, presumably starting with the specific ones we already found by numerical continuation. This could possibly be done by setting up Lin's method as a theoretical tool (rather than a numerical one) \cite{PARKER2021132890,KNOBLOCH20142984}. Moreover, 
our computations based on boundary-value problem formulations come with error bounds; hence, the existence of any connecting orbit we showed here can be proved, in principle, in a computer-assisted way \cite{Wilczak2003,Kapela2021}. Another issue is the stability of the different solitons as solutions of the PDE. Only the primary single-hump soliton is stable \cite{tam2018solitary,GDK}, while all the other solitons we discovered appear to be unstable (but some are only weakly unstable \cite{PhysRevA.103.063514}). Establishing this observation rigorously remains a considerable challenge for future work and will require state-of-the art techniques, such as the computation of Evans functions \cite{barker2018evans}. 

Finally, localized structures in other spatially extended systems may be studied and explored, in the same spirit, with the numerical continuation approach employed here. 
In particular, recent experiments \cite{PhysRevResearch.3.013166} have shown the feasibilities of creating waveguides with higher even-order dispersions, which are described by further generalizations of the NLSE with corresponding non-zero dispersion terms of order six, eight, ten, or even higher. Other examples of interesting PDEs in this context are the widely studied class of the NLSE describing solitary wave formation in inhomogeneous media \cite{Kominis:08,PhysRevA.87.063849,PhysRevE.100.052209,KOMINIS2019222}, and the Lugiato-Lefever equation \cite{articleParra}, which both are known to feature homoclinic snaking.

\section*{Acknowledgments}
A. Giraldo was supported by KIAS Individual Grant No. CG086101 at the Korea Institute for Advanced Study.


\section*{Data availability}
The data that support the findings of this study are available from the corresponding author upon reasonable request.

\bibliographystyle{plain}

\bibliography{bgbk_perGNLSE_references} 

\end{document}